\documentclass[aps,pra,twocolumn,showpacs]{revtex4-1}

\usepackage{graphicx}
\usepackage{amsmath,amsfonts,amssymb,amscd,amsthm,xspace} 
\usepackage{makeidx}
\usepackage{color}    
\usepackage{amssymb,verbatim}
\usepackage{amsmath}
\usepackage{txfonts}
\usepackage{bbm}
\usepackage{wrapfig}
\usepackage{soul}
\usepackage{calligra}
\usepackage{appendix}

\newcommand{\tr}{\mbox{tr}}
\newcommand{\bra}[1]{ \langle #1 |}
\newcommand{\ket}[1]{| #1 \rangle}

\usepackage[breaklinks=true,colorlinks=true,linkcolor=blue,urlcolor=blue,citecolor=blue]{hyperref}

\begin{document}


\title{Macroscopic superpositions and gravimetry with quantum magnetomechanics}


\author{M.T. Johnsson, G.K. Brennen, and J. Twamley}
\affiliation{Centre for Engineered Quantum Systems, Department of Physics and Astronomy, Macquarie University, Sydney, NSW 2109, Australia}


\date{\today}

\begin{abstract}
We utilise a magneto-mechanical levitated massive resonator in the quantum regime to prepare highly macroscopic quantum superposition states. Using these macroscopic superpositions we present a novel interferometry protocol to perform absolute gravimetry with a sensitivity that exceeds state of the art atom-interferometric and corner-cube gravimeters by a factor of 20. In addition, our scheme allows probing the gravitational field on a length scale eight orders of magnitude smaller than other methods.
\end{abstract}


\maketitle

\section{Introduction}
Absolute gravimetry measures the local acceleration due to gravity on a test body. Precision gravimetry has numerous applications ranging from metrology, geophysics,  geodesy and inertial sensing \cite{Reynolds:2011ub, deAngelis:2008jla,Krynski:2012ti}, through to precision measurements of the fine-structure constant in Quantum Electrodynamics (QED) \cite{Bouchendira:2011wk}, the gravitational constant \cite{Bertoldi:963356, 2007Sci...315...74F,Rosi:2015kv},  testing alternative theories of gravity and quantum gravity \cite{Dimopoulos:2007wv,Albrecht:2014ex,Kafri:2014cx}, and potentially searching for gravitational waves \cite{Tino:2007is, Tino:2011bx}. One of the prototypical standard instruments used, the Scintrex FG-5, is based on a free falling corner cube combined with a Mach-Zehnder interferometer and atomic clock   \cite{1995Metro..32..159N}. By utilising  advanced isolation techniques this instrument can achieve an absolute gravimetry precision of ${\sim}\,15\, \mu{}\text{Gal}\,{\text{Hz}}^{-1/2}$,  ($1 \, \mu {\mathrm{Gal}} = 10^{-8}$\,ms$^{-2} \sim 10^{-9}\, g$). However atom based interferometers have been demonstrated to exceed this with initial experiments yielding a precision of ${\sim}\, 8\, \mu{}\text{Gal}\,{\text{Hz}}^{-1/2}$ \cite{Muller:2008ik}, and more recently, using an optimised active isolation system, a precision of ${\sim}\, 4.2\, \mu{}\text{Gal}\,{\text{Hz}}^{-1/2}$ \cite{Hu:2013kq}. Proposals using large area atomic interferometers  or by extending the duration of the free-fall via micro-gravity/space based setups, predict that atomic gravimeters might reach precisions of ${\sim}\, 10^{-4}\, \mu{}\text{Gal}\,{\text{Hz}}^{-1/2}$ \cite{Chiow:2011eu,Tino:2013ii}, but so far the best precision demonstrated is that achieved by Hu {\it et al.} \cite{Hu:2013kq}. Both of these techniques require long (${\sim}\, 1$m), fall drops and thus they give a spatially averaged result. 

In this work we describe how one can perform absolute gravimetry using a quantum  magnetomechanical system consisting of a magnetically trapped superconducting massive mechanical resonator in vacuum whose motion is controlled and measured by a nearby RF-SQUID or flux qubit. By driving the mechanical massive resonator to be in a macroscopic superposition of two different heights we are able to execute an interferometry protocol which has the potential to achieve a gravimetry precision of ${\sim}\, 0.22\, \mu{}\text{Gal}\,{\text{Hz}}^{-1/2}$, with a spatial resolution of a few nanometres. Furthermore, this value is limited only by the coherence time of the flux qubit, offering the possibility of significant improvements to the precision in the future.

Our scheme is based on engineering large spatial superpositions of a massive object. Generating macroscopic quantum superpositions has been a much sought after goal both from the viewpoint of studying  fundamental issues relating to the classical/quantum boundary  but also towards using such superpositions for enhanced sensing. Examples include experiments indicating the quantum matter-wave nature of individual high-weight organic molecules and quantum motional oscillations of membrane based optomechanical systems  \cite{Arndt:2014bp}. Quantum magnetomechanics (as opposed to optomechanics), uses magnetic forces as opposed to light forces, towards engineering the quantum motion of systems. One significant advantage of the former is the greatly reduced motional noise with  passive magnetic forces as compared with scattered light induced noise in optomechanical systems. The reduced noise in magnetomechanical systems uniquely permits the engineering of ultra-high (${\sim}\, 10^9$), motional $Q$-factors in magnetically levitated massive resonators \cite{Cirio:2012gk, RomeroIsart:2012hg}, while incorporating magnetostrictive elements one can design hybrid quantum systems to couple microwave and optical quantum signals \cite{Xia:2014ku}.

In the following we first describe the model magnetomechanical quantum system consisting of a superconducting ring stably trapped and levitated in vacuum within the inhomogeneous magnetic field generated by a small magnetic sphere. The motion of this ring can be cooled \cite{Cirio:2012gk}, and  coherently controlled via an inductively coupled nearby superconducting flux qubit which is controlled via a superconducting electrical circuit. We next introduce a metrological interferometry protocol where we drive, via the flux qubit, the generation of counter-oscillating  vertical motional cat states of the trapped massive ring. We show that by measuring the state of the qubit at the end of the interferometry dynamics one can obtain a direct measurement of the accrued phase shift and from this, the local acceleration due to gravity. To achieve a high precision and dynamic range for this metrology we tune the spatial scale of these counter-oscillating superposition states via adjusting the currents in the flux qubit. Through this we are able to propose a protocol that can estimate the local acceleration due to gravity over a large dynamic range and with very high precision. We estimate the ultimate possible precision we could hope to achieve in the near future assuming progress in extending  the coherence times of flux qubits. We then discuss various sources of noise and decoherence and discuss briefly the possibilities for an experimental implementation.

\section{Model}
\label{Model}
We first discuss in more detail the physical setup for our magnetomechanical gravimeter. We study the three dimensional (3D) trapping of the ring, the oscillation frequency and stability, the inductive coupling between the ring resonator and the flux qubit, the desired characteristics of the permanent magnet providing the inhomogeneous trapping flux, the materials and dimensions for the ring resonator, and the coherence properties and operation of the flux qubit. This will lead on to our description of the actual metrology protocol in the following section. 

\subsection{Setup}
\begin{figure}
\includegraphics[ width=7.5cm, height=8.8cm]{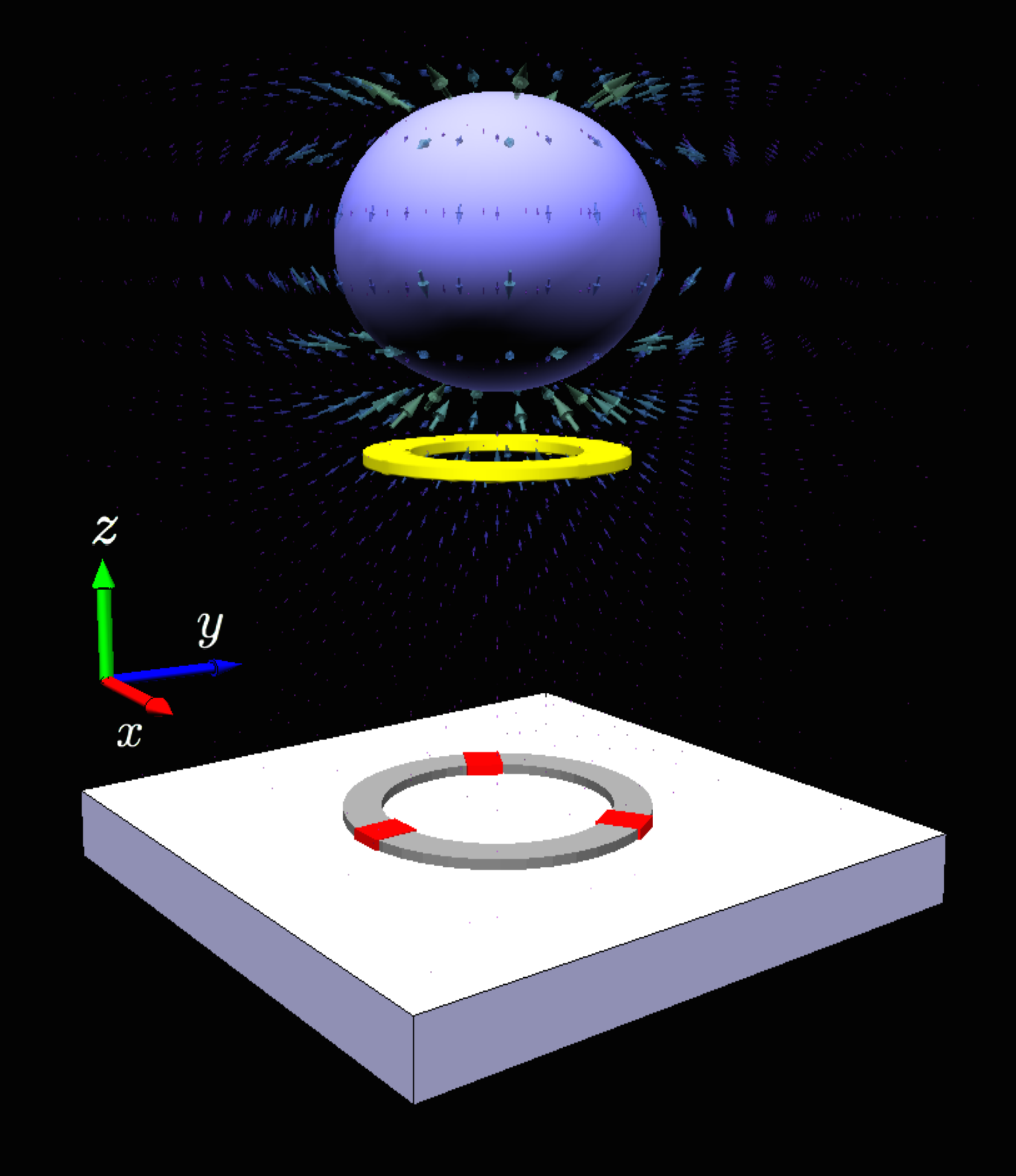}
\caption{Schematic of the gravimeter setup. A small permanent magnet (blue sphere) with uniform magnetization in the upwards vertical direction produces a strong spatially inhomogeneous magnetic flux (shown as a vector field). A small superconducting ring (resonator - yellow), is trapped via the Meissner effect below the sphere.  Currents in a flux qubit (grey) on a substrate (white) couple inductively to the motion of the trapped ring and the qubit can be used either to cool or coherently control the motion of the ring. The state of the qubit is readout using a DC SQUID - not shown.}
\label{Fig1}
\end{figure}

As depicted in Fig \ref{Fig1}a, we consider a small permanent magnet whose purpose is to provide a highly spatially inhomogeneous magnetic field. We choose a sphere, radius $R_s$, volume $V$, to have uniform magnetization ${\bf\cal M}={\cal M}\hat{z}$, where  $\hat{z}$  is oriented vertically upwards. We have found that 3D magnetomechanical trapping can occur with various shaped magnets such as cones and spheres, but we choose the sphere for simplicity as the resulting fields, fluxes and potentials can be derived analytically.

We consider this magnet to be rigidly fixed while the underlying flux-pinned ring resonator can oscillate freely. Trapped a distance $z_{\mathrm{eq}}$ below the center of the sphere is a ring (which we will denote as the resonator), of radius $R_r$ of superconducting wire of thickness $2a$ (circular cross section), lying in the $\hat{x}-\hat{y}$ plane with self-inductance $L_r$.  Small spatial oscillations of this resonator will lead to small changes in the relative position of the centre of the ring with respect to the centre of the sphere.   As the resonator moves through the inhomogeneous magnetic field the enclosed flux due to the sphere will change and the Meissner effect will cause supercurrents to be generated within the ring to maintain the overall enclosed flux $\Phi$, constant in time. The magnetic fields generated by these supercurrents will interact with the sphere's magnetic fields causing a mechanical restoring force on the resonator leading to trapping of the resonator in all three directions \cite{Cirio:2012gk}.

Located below the resonator is a superconducting flux qubit. The flux qubit generates counterpropagating supercurrents which can be in quantum superposition. The currents circulating in the flux qubit generate a magnetic field and these couple via mutual inductance $M_{rq}$ to the currents flowing in the resonator. With this coupling one can use the flux qubit to  cool the motional state of the resonator \cite{Cirio:2012gk}, but in addition one can use the flux qubit to coherently drive/control the motion of the resonator. We will use this latter capability to perform the interferometry protocol. The precision in the gravimetry protocol is directly related to how strong we can engineer this resonator-qubit coupling and this coupling decreases as the resonator-qubit separation increases.

Provided the resonator is initially cooled to superconducting temperatures at some distance from the sphere, it will have zero magnetic flux threading it, and no persistent supercurrents. When it is moved into place below the sphere, supercurrents will be induced to ensure that it continues to have zero flux threading it, even though it is directly below the sphere. We will work with a resonator that is the same size or larger than the flux qubit as well as being in very close proximity to it, ensuring that the qubit is shielded from any flux noise arising in the magnet.

\subsection{Oscillation frequencies and stability}

We now analyse the  mechanical trapping  of the resonator in 3D and estimate the resonator's vertical trapping frequency and discuss the transverse trapping dynamics. We assume the magnetisation of the sphere is along the $\hat{z}$ axis, i.e. ${\boldsymbol{\cal{M}}} = {\cal M} \hat{z}$. In cylindrical coordinates $(\rho, \phi, z)$, with the origin at the centre of the sphere, the vector potential for such a homogeneously magnetized sphere is 
${\mathbf A} (\mathbf{r}) = \mu_0 {\cal M} V \rho\,(\rho^2 +z^2)^{-3/2}\hat{\phi}/4 \pi\, $,
where $V$ is the sphere volume. 
Since ${\mathbf B} = \nabla \times {\mathbf A} $ we have
${\mathbf B} (\mathbf{r}) = \mu_0 {\cal M} V (\rho^2 +z^2)^{-5/2}(3z\,\rho,0,2\,z^2-\rho^2)/4\pi$, in cylindrical coordinates.
To calculate the vertical trapping frequency $\omega$, we require the force exerted on the resonator ring as a function of displacement from the equilibrium point. This force arises due to the fact that when the ring moves, the magnetic flux threading it flux will change. As flux lines can't pass through the superconducting ring, however, a current arises in the ring to restore the flux, and this current gives rise to a Lorentz force. 

Since the resonator ring is horizontal, the flux through it will be given by the line integral
\begin{equation}
\Phi = \oint {\mathbf A} (\mathbf{r}) \cdot \mbox{d} \mathbf{r} = \frac{\mu_0 {\cal M} V R_r^2  }{2(R_r^2 + z^2)^{3/2}}\;,\label{eqFluxViaPhiIntegral}
\end{equation}
where $R_r$ is the radius of the resonator ring. Changes in current in the resonator as it moves vertically are related to changes in flux via 
\begin{equation}
\frac{dI_r}{dz} = -\frac{1}{L_r} \frac{d \Phi} {dz}\;,
\label{eqdIdz}
\end{equation}
where $L_r$ is the self inductance of the ring and is given by
$L_r = \mu_0 R_r (\ln [8R_r / a] - 2)$, 
where the ``$-2$'' factor indicates we assuming all the current is on the surface of the resonator.

Taking the equilibrium vertical position point $z=z_{\mathrm{eq}}$, then for small displacements the current in the resonator is
$I_r = (z-z_{\mathrm{eq}})\cdot \left. dI_r/dz\right|_{z=z_{\mathrm{eq}}}$ 
where, using (\ref{eqdIdz}), 
\begin{equation}
\left.\frac{d I_r} {dz}\right|_{z=z_{\mathrm{eq}}} = \frac{3 \mu_0 {\cal M} V R_r^2 z_{\mathrm{eq}} }{2L(R_r^2 +z_{\mathrm{eq}}^2)^{5/2}}.
\label{eqDIdz1}
\end{equation}
The Lorentz force from the current, magnitude $I_r$, flowing through a small element $\mbox{d} {\mathbf l}$ of the wire is given by $\mbox{d}{\mathbf F} = (I_r \, \mbox{d} {\mathbf l}) \times {\mathbf B}$.
Assuming the resonator is circular, sitting horizontally, and is co-axial with the $\hat{z}-$axis, the line element $\mbox{d} {\mathbf l}$ will always be perpendicular to ${\mathbf B} = B_{\mathrm{radial}} \hat{ \rho}+B_{\mathrm{axial}}\hat{z}$. Hence the vertical force on the resonator for small vertical displacements from equilibrium $z-z_{\mathrm{eq}}$, is
\begin{equation}
F_z = - I_r \int_0^{2\pi} B_{\mathrm{radial}} R_r d\phi 
    =  -\frac{9 \mu_0^2 {\cal M}^2 V^2 R_r^4 z_{\mathrm{eq}}^2 (z-z_{\mathrm{eq}})}{4 L_r (R_r^2 +z_{\mathrm{eq}}^2)^5}.
\label {eqFz}
\end{equation}
Finally, the equation of motion in the $z$ direction is
\begin{equation}
\frac{d^2 z} {dt^2} = \frac{F_z}{ m} = -\omega^2 (z-z_{\mathrm{eq}}) \label{eqEOMomega}
\end{equation}
for small displacements, providing an harmonic restoring force. Comparing (\ref{eqFz}) and (\ref{eqEOMomega}) we find the vertical oscillation frequency
\begin{equation}
\omega = \frac{3 \mu_0 {\cal M} V R_r^2 z_{\mathrm{eq}}}{2\sqrt{m L_r (R_r^2+z_{\mathrm{eq}}^2)^5}}\;.
\label{eqDefnForomega}
\end{equation}
We also need to consider transverse trapping and oscillations firstly to establish that the resonator is indeed trapped in all three directions, and secondly to determine if there is any coupling between the vertical and horizontal motions. If this coupling exists then by cooling the vertical motion one cools the entire motion of the resonator, but such couplings can also lead to unwanted energy leakage from the coherent vertical dynamics to the transverse modes, leading to decoherence of our vertical superposition states.

As shown in Appendix \ref{secTransverseTrapping}, to lowest order the trapping potential is given by
\begin{equation} 
V = \frac{1}{2} m \omega^2 z^2 + \frac{1}{3} \gamma (x^2 + y^2) z + \frac{1}{4} \beta (x^4 + y^4) ,
\end{equation}
which describes a type of cross-mode coupling. For parameters described in Appendix \ref{secSymbols} we find $(m \omega^2 /2, \gamma, \beta)= (1.73 \times 10^{-2}$\,J, $1.98 \times 10^3$\,Jm$^{-1}$, $2.65 \times 10^8$\,Jm$^{-2}$).

The horizontal trapping at equilibrium ($z=z_{\mathrm{eq}}$) exhibits {\emph {extremely}} slow oscillations. The period is amplitude-dependent, with higher amplitudes having shorter periods, but even with an unrealistically large amplitude of a 10\,$\mu$m the period is ${\sim}\,$50 seconds. This  means that the horizontal dynamics are essentially frozen out when compared with the fast vertical dynamics of the resonator. 


\subsection{Inductive coupling to the qubit}

The coupling between the resonator and the qubit is determined by the mutual inductance between the currents flowing in the qubit and the small currents flowing in the resonator, the latter being dependent on the vertical position of the resonator. This coupling is of the form
$\hat{H}_{\mathrm{coupling}} = \hbar \lambda (\hat{a} + \hat{a}^{\dagger}) \hat{\sigma}^z/2$,
where $\hat{\sigma}^z$ describes the direction of the current in the qubit, $\hat{a}$, the annihilation operator for vertically trapped motional resonator phonons, and the coupling strength $\lambda$ is defined as
\begin{equation}
\lambda = \sqrt{\frac{2}{m \hbar \omega}} \, M_{rq} \, \frac{dI_r}{dz}|_{z=z_{\mathrm{eq}}} \,\, I_q, \label{eqLambdaDefn}
\end{equation}
where $M_{rq}$ is the mutual inductance between the resonator and the qubit, $I_q$ is the current in the qubit, $dI_r/dz$ describes how the induced current in the resonator changes with respect to its vertical displacement from the equilibrium point $z_{\mathrm{eq}}$, and $m$ is the mass of the resonator. 

The mutual inductance between two parallel  rings, one co-axial above the other, radii $R_r$ and $R_q$, and co-axial separation $d$, is
\begin{equation}
M_{rq} = \mu_0 \sqrt{\frac{4 R_r R_q}{\eta(d)}} \left( \frac{\mbox{K} \left(\eta(d) \right)}{1+\beta(d)}
                     - \mbox{E}\left( \eta(d)\right) \right),
\label{eqMutualInductance}
\end{equation}
where $\eta(d)\equiv2\beta(d)/(1+\beta(d))$, $\beta(d) = 2 R_r R_q/(R_r^2 +R_q^2 +d^2)$, $\mbox{K}$ is the elliptic integral of the first kind and $\mbox{E}$ is the elliptic integral of the second kind.
Although $M_{rq}$ scales as $1/d^3$ for large separations, we will work in the regime where $d< R_r, R_q$, so as to maximise the inductive coupling strength. 
Using Eqs.~(\ref{eqDIdz1}), (\ref{eqDefnForomega}), (\ref{eqLambdaDefn}) and (\ref{eqMutualInductance}) we obtain
\begin{equation}
\lambda = \sqrt{ \frac{3 {\cal M} V \mu_0 }{\hbar}} \left[ \frac{R_r^4 \, z_{\mathrm{eq}}^2}{m L_r^3 \,(R_r^2 +z_{\mathrm{eq}}^2)^5} \right] ^{1/4} M_{r,q} \, I_q\;,
\label{eqLambdaDefinition}
\end{equation}
where $z_{\mathrm{eq}}$ is the vertical distance from the centre of the sphere to the resonator. The coupling $\lambda$ is proportional to the square root of the magnetisation and directly proportional to the current in the qubit.

\begin{figure}
\includegraphics[width=8cm]{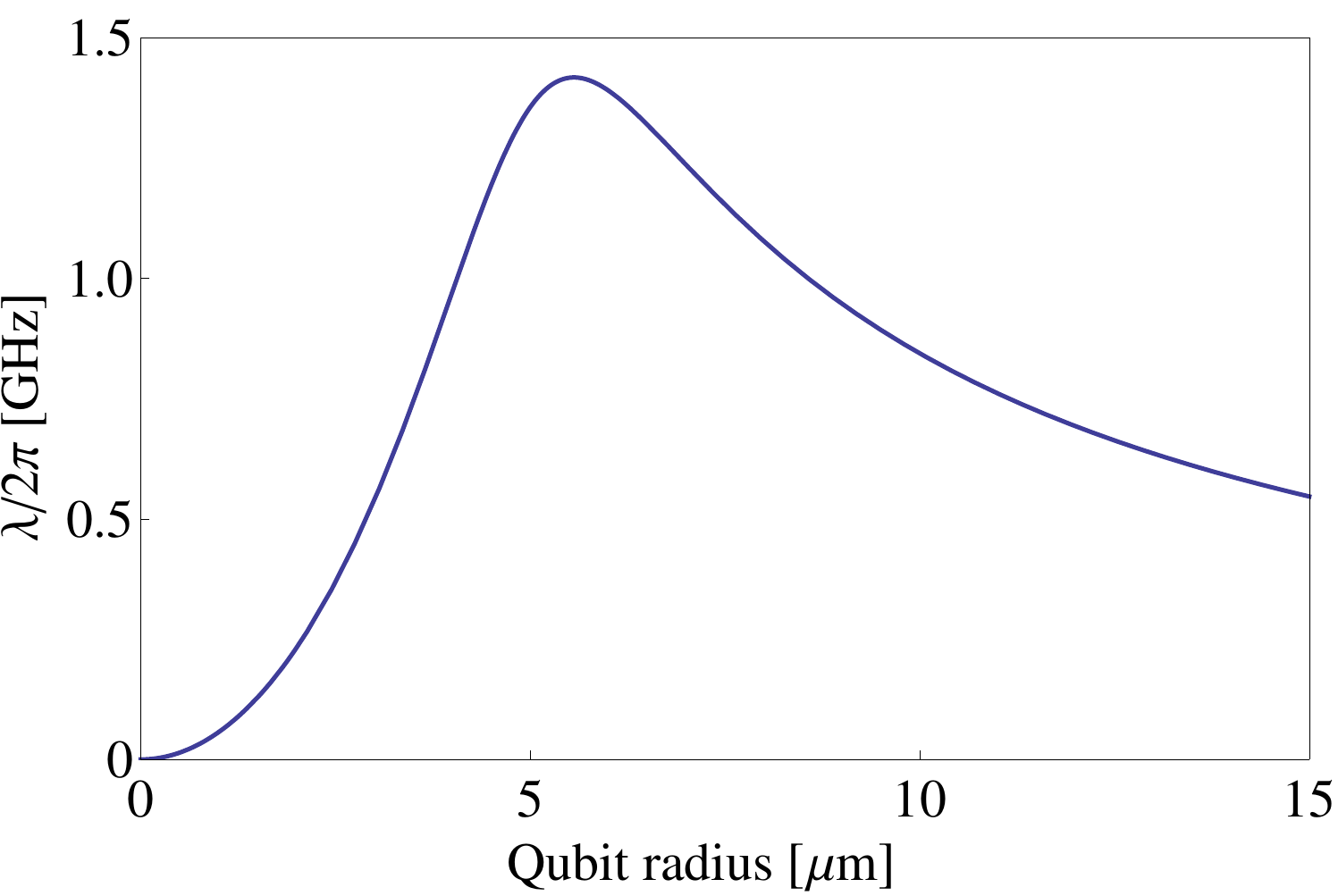}
\caption{Value of the inductive coupling strength $\lambda$ between the resonator and qubit for a fixed resonator radius of $5\,\mu$m and a fixed sphere radius of $10\,\mu$m as we vary the qubit radius. Maximal $\lambda$ requires $R_r=R_q$. System parameters as given in Appendix \ref{secSymbols}.}
\label{figLambdaRrAgainstRq}
\end{figure}

Maximising $\lambda$ requires the size of the qubit and resonator to be near identical $R_q{\simeq}\, R_r$. Assuming both are circular we obtain the results shown in Figure \ref{figLambdaRrAgainstRq}. In addition, provided that we have control over the radius of the magnetic sphere, the maximal value for $\lambda$ is obtained when the radius of the sphere is twice the radius of the qubit and resonator, i.e. $R_{sphere} = 2 R_{q} = 2R_r$.

\subsection{Materials for the setup}
To reduce  damping loss of the resonator's motion due to induced eddy currents in the spherical magnet one can consider the magnet to be made from a suitable magnetic insulator, e.g. Yttrium-Iron-Garnet (YIG), which possesses a saturation magnetization of $\mu_0 {\cal{M}} {\sim}\, 0.17$\,T. Due to the resonator's close proximity to the magnetised sphere and the fact that inductive coupling to the qubit results in large current densities, we require it to be composed of a superconductor with a high critical current and a high critical magnetic field. Further, to avoid decoherence due to flux pin dragging of the sphere's magnetic field as the resonator oscillates \cite{Niemetz:2000wk}, we require a Type-I rather than a Type-II superconductor. For these reasons we choose lead, which has a critical temperature $T_c\,{\sim}\, 7$\,K and critical field $H_c\, {\sim}\, 0.08$\,T, and limit the magnetisation of the sphere to this field strength. Qubits are typically fashioned from either Aluminium or Niobium with the latter having the advantage of a higher critical magnetic field $H_c{\sim}\, 0.83$\,T thus permitting the qubit to remain superconducting in the presence of the magnetic sphere.

\subsection{Qubit subsystem}
The superconducting qubit subsystem we use is essentially a flux qubit. This is a superconducting loop containing a Josephson-junction (JJ). The flux qubit is driven by a magnetic flux which is generated by an external nearby circuit and recent versions of the flux qubit involve two identical JJs and a smaller JJ. This three junction design allows for large persistent currents with a small geometrical size (and thus inductance) of the superconducting loop and results in an operation which is less sensitive to noise.

\begin{figure}
\includegraphics[width=8cm]{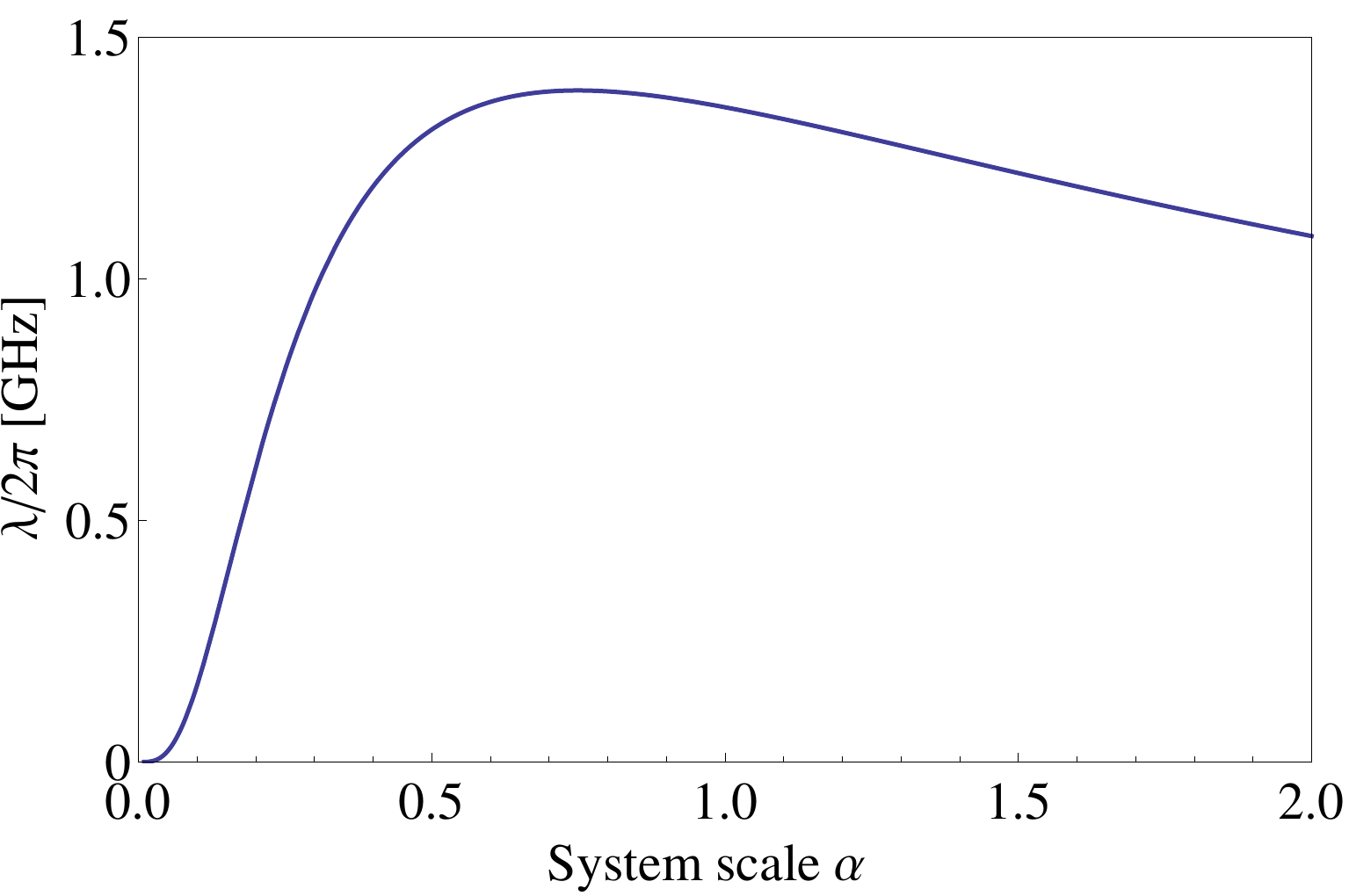}
\caption{Behaviour of the inductive coupling strength between the motion of the resonator and qubit $\lambda$, as a function of overall system size, assuming the largest current that can be achieved using a flux qubit of a specific size. We take $R_r = R_q = 5 \alpha \, \mu$m, $R_{\mathrm{sphere}}=10 \alpha \, \mu$m, $a=1.0 \alpha\,\mu$m. Static parameters: $d=2\,\mu$m, $r_0 = 1\,\mu$m (see Table \ref{default}).}
\label{fig_lambda_vs_system_size_optimized_for_flux_qubit}
\end{figure}

Recalling the operation of an RF-SQUID (similarly a flux qubit) \cite{Barone:101094},
one applies a controlled external magnetic flux bias to yield an effective double-welled potential for the Hamiltonian of the qubit whose lowest energy symmetric/antisymmetric wave functions act as a two level system. These states correspond to oppositely circulating currents in the qubit loop and are split in energy depending on the height of the double well tunnel barrier. This splitting is quantified by the quantity $\upsilon = L / L_J - 1$, where $L$ is the geometric inductance of the qubit, and $L_J$ is the Josephson inductance. Roughly speaking, $L_J$ is controlled by the size and thickness of the junction, while $L$ is set by the size and shape of the qubit loop. The energy splitting between the two levels increases as $\upsilon \rightarrow 0$, which means we need $L\,{\sim}\, L_J$ in order to operate in a regime where the two qubit levels are sufficiently split in energy.
Since the inductance of a circular wire loop  of radius $R$ is roughly proportional to $R \ln (R)$, one cannot engineer very large qubit loops while still retaining the relation $L \,{\sim}\, L_J$. As the inductive coupling also is proportional to the current in the qubit we require this also to be large but this is in conflict with large loop area as $I_{max}\, {\sim}\, \Phi_0/2 L$. Thus there is a trade-off between the maximal  current and physical size of the qubit --- each contributing to the overall inductive coupling. For definiteness one can study the resonator-qubit inductive coupling strength $\lambda$ as one scales up the physical size of the qubit/resonator/magnet (see Figure~\ref{fig_lambda_vs_system_size_optimized_for_flux_qubit}), and curiously the optimal scale yielding the largest coupling strength is achieved when the flux qubit circular loop is quite small with radius ${\sim}\,$5--10$\,\mu$m. 

\subsection{Cooling to the ground state}

In order to put our resonator into a cat state and use it as a gravimeter, it is necessary to ensure that we can begin with it in the motional ground state. This in turn requires that we have mechanism to cool it from its initial non-equilibrium state to the ground state by removing energy.

Details of the cooling scheme we use can be found in Refs. \cite{Cirio:2012gk,Rabl:2010cm}, which we will briefly summarize here. We cool by coupling a two level system (the qubit) to the resonator, with the qubit coupled to a bosonic thermal bath. The Hamiltonian for the coupled system is
\begin{equation}
\hat{H} = -\frac{\hbar \delta}{2} \hat{\sigma}^z + \frac{\hbar \Omega}{2} \hat{\sigma}^x 
           + \hbar \omega \hat{a}^{\dagger} \hat{a} + \frac{\hbar \lambda}{2} (\hat{a} + \hat{a}^{\dagger}) \hat{\sigma}^z
\label{eqCoolingH}
\end{equation}
where $\Omega$ is the Rabi frequency with which we drive the qubit, $\delta$ is the detuning of the driving field from resonance with the qubit frequency splitting $\omega_q$, $\hat{a}$ is the annihilation operator for the resonator oscillation modes and the $\hat{\sigma}^{x,z}$ are the standard spin-1/2 Pauli operators. 

The open systems dynamics of the qubit-resonator system is described in Appendix \ref{Decoherence} and is characterised by $\Gamma$ and $\Gamma_{\perp}$, the amplitude damping rates of the resonator and qubit respectively. The initial state of the resonator is modelled as a coherent state with amplitude $\alpha=\sqrt{N_{\rm th}}$ where the initial occupation number is $N_{\rm th}=(e^{\hbar\omega/k_B T_r}-1)^{-1}$ with $T_r$ an effective bath temperature for the environment of the resonator.  In the limit where $\lambda \ll \Gamma_{\perp}, \omega$, the final phonon occupation number for the resonator, $n_f$, is given by 
\begin{equation}
n_f = N_{\mathrm{th}} [ \zeta + (1-\zeta)/(1+\zeta \exp[I_1/(N_{\mathrm{th}} \zeta (\lambda/\omega)^2)])].
\end{equation}
Here $\zeta = \Gamma/\Gamma_c(0)$ and the renormalized cooling rate is $\Gamma_c(\alpha) = i\lambda (\vec{S}_1^z/\alpha -\vec{S}^z_{-1}/\alpha^*)$, with $I_1 = 2 \int^{\infty}_0 d\alpha \, \alpha \tilde{\Gamma}_c(\alpha \omega/\lambda)$ and $\tilde{\Gamma}_c = \Gamma_c(\alpha)/\Gamma_c(0)$. 
The qubit polarization Fourier components, $\vec{S}_1^z$ and $\vec{S}_{-1}^z$, are given by the solutions to the Bloch equations for the qubit. In the Lamb-Dicke regime ($\lambda \sqrt{N_{\mathrm{th}}+1/2} \ll \Gamma_{\perp}, \omega$) one can obtain an effective master equation for the resonator after tracing out the qubit. This gives a new effective resonator damping rate $\Gamma_{\rm cool} = \Gamma_c + \Gamma$ with $\Gamma_c = S(\omega) - S(-\omega)$ where $S(\nu)$ denotes the qubit fluctuation spectrum and is given by
\begin{equation}
S(\nu) = \frac{\lambda^2}{2} {\mathrm{Re}} \int^{\infty}_0 e^{i\nu t} dt [\langle \hat{\sigma}^z (t) \hat{\sigma}^z(0) \rangle_0 - \langle \hat{\sigma}^z(0) \rangle_0 ^2],
\end{equation}
where $\langle \cdot \rangle_0$ denotes the steady state expectation. The resulting steady state phonon occupation of the resonator in the Lamb-Dicke regime is  \cite{Rabl:2010cm}
\begin{equation}
n_{LD} = \Gamma N_{\mathrm{th}}/\Gamma_c + N_0,
\label{eqnLD}
\end{equation}
where $N_0 = S(-\omega)/\Gamma_c$.

\begin{figure}
\includegraphics[width=9cm]{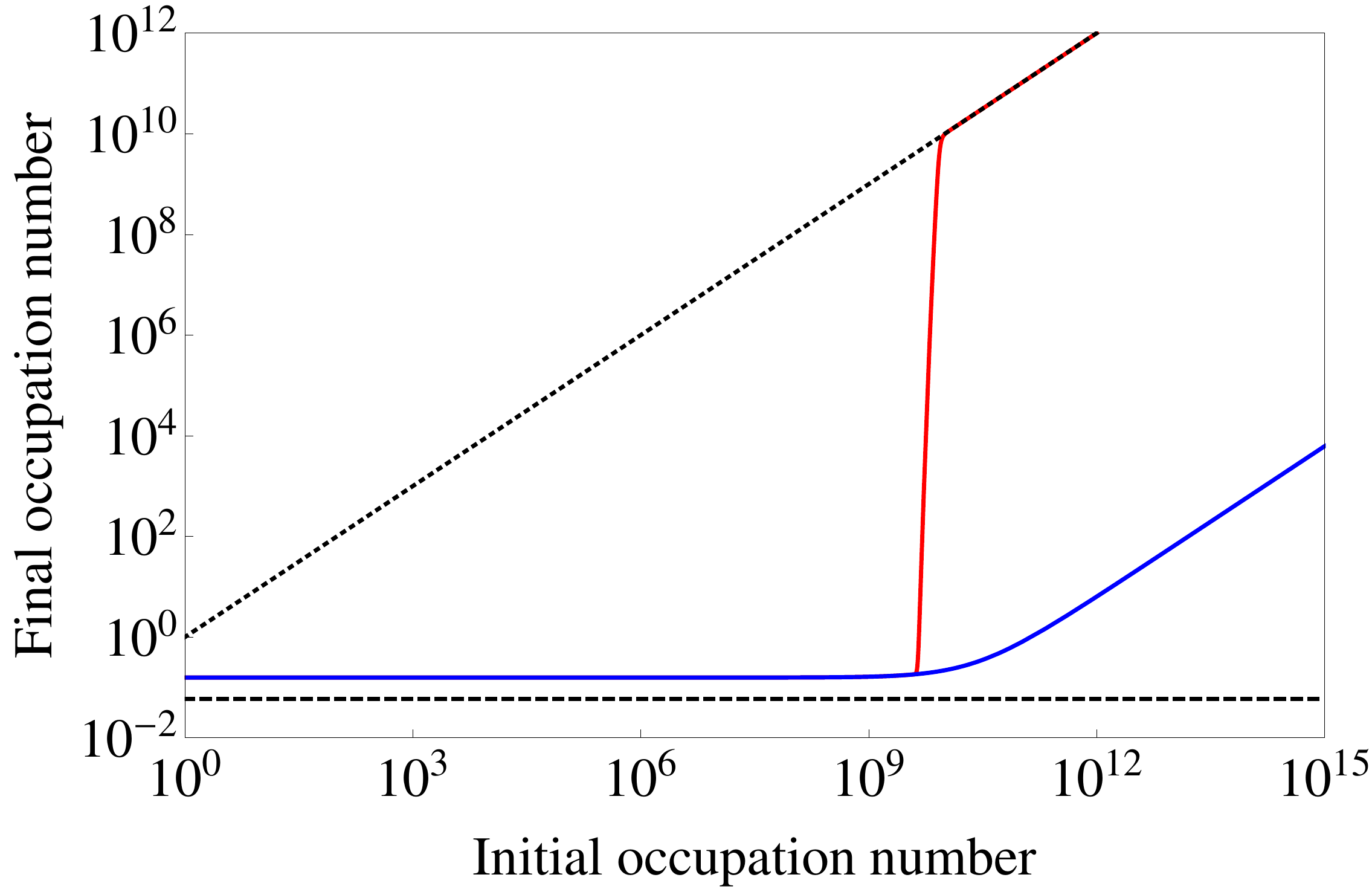}
\caption{Cooling performance of our system. Horizontal dashed black line shows the phonon occupation of the qubit in a temperature bath of $T_q=100$\,mK, solid blue line shows the final phonon occupation number of the resonator after cooling assuming the Lamb-Dicke regime is valid, and the solid red line shows the exact cooling solution (note blue and red lines overlap for low initial occupation numbers). The diagonal dotted black line shows the line where initial and final occupation numbers are the same; any part of the blue or red curves below this represents cooling. The plot shows that even with initial phonon occupation numbers as high as ${\sim}\,10^9$ we can cool the resonator to the ground state, with an average final occupation number of 0.16.
We have assumed the standard resonator parameters given in Table \ref{default}, and take $\Omega=\omega/2$, $\delta = -\sqrt{\omega^2 - \Omega^2}$, $\lambda/2\pi = 10^4$\,Hz, decoherence times $T_1 = T_2 = 70\,\mu$s, $\omega_q/2\pi = 6$\,GHz, and $\Gamma=2.70 \times 10^{-8}$\,Hz. See Appendix \ref{secSymbols} for the definitions of system parameters.}
\label{fig_cooling_performance}
\end{figure}

In Figure~\ref{fig_cooling_performance} we plot the performance of this cooling scheme for our system, showing both the full cooling solution and a simplified cooling solution that makes the assumption that we are always in the Lamb-Dicke regime, i.e. Eq.~(\ref{eqnLD}) holds for all initial resonator temperatures. The plot shows that even with initial phonon occupation numbers as high as ${\sim}\,10^9$ we can cool the resonator to the ground state, with an average final occupation number of 0.16.

The timescale governing the cooling is given by the effective resonator cooling rate $\Gamma_{\rm cool}$. For our system, using parameters given in the caption of Figure~\ref{fig_cooling_performance}, we obtain $\Gamma_{\rm cool} = 27\,$kHz. We note this cooling rate scales as $\lambda^2$, and we have chosen a very conservative coupling rate of $\lambda = 10$\,kHz. As our system is capable of coupling strengths of up to $\lambda \, {\sim}\, 1$\,GHz, the cooling can be made much faster if required.
 
\section{Metrology protocol}

\subsection{Description}

We now describe in general terms the operation of the gravimetry protocol. We then go into more details regarding the precision one might expect using a simple interferometric protocol. Before starting the protocol one must prepare the resonator in a levitated 3D trapped state and in the ground state of vertical motion as described above. We arrange, via tuning the frequency of the qubit for instance, to turn off the resonator-qubit coupling and to initialize the qubit in the state $|+_x\rangle=(\ket{1}+\ket{-1})/\sqrt{2}$. Next the resonator-qubit coupling is turned on. Notice that the effect of the qubit in $\pm 1$ eigenstates of $\hat{\sigma}^z$, $|\pm 1\rangle$, is to apply slightly different constant forces on the resonator in the vertical direction. These forces cause slight displacements in the trapping potential of the resonator providing for spatial superposition states which evolve in state dependent traps displaced from each other in the vertical direction.  We then let the resonator evolve in these state dependent traps and after a specific duration the resonator will return to its initial height (which we denote as ``one slosh''). However because of the slight difference in heights of the two traps a phase difference will accrue and when the resonator returns to its original height one will obtain constructive/destructive interference. This interference can be probed by again quickly turning off the resonator-qubit coupling and by measuring the qubit along the $\hat{x}-$axis of its Bloch sphere. We will see that the phase difference will be directly proportional to the absolute acceleration due to gravity and we conclude with a rough estimate of the precision one might expect under a naive interferometric estimation protocol. In the following section we detail a more sophisticated estimation protocol that can yield far greater precision and dynamic range in the gravimetry.

\begin{figure*}
\centering
\includegraphics[width=16cm]{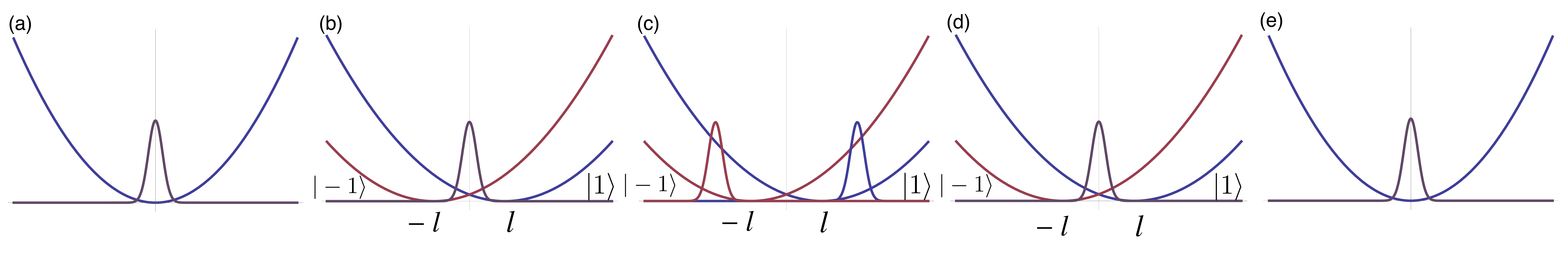}
\caption{Illustration of one stage of the splitting protocol for gravimetry. Shown are the spin dependent harmonic trapping potentials along the $\hat{z}$ direction and the one dimensional  spatial wave function for the resonator as a function of $z$.  This protocol is repeated $M(K,k)$ times for each value of the qubit resonator coupling $\lambda_k=2^k\lambda_0$, for $k=0,\ldots K$ where $\lambda_0$ is a minimal value of the coupling.  (a) At $t<0$ the resonator is prepared in the ground motional state with an rms width $z_0$ and frequency $\omega$. The qubit is not coupled to the resonator $(\lambda_k=0)$. (b) At time $t=0$ the qubit in prepared in the superposition state $\ket{+_x}$ and the interaction $\lambda_k$ is turned on. The trapping potential is now state dependent with minima located at $\pm l=\pm \lambda_k z_0/\omega$. (c) After half an oscillation period, the state dependent motional wave packets are maximally separated by a distance $4l$. (d) After a full oscillation period, the wave packets recombine, localised at the origin with the accumulated gravitationally induced phase $\phi_k$ mapped onto the qubit. (e) The interaction is turned off and the qubit is measured.  For $\lceil M(K,k)\rceil$ rounds the qubit is measured in the basis $\{\ket{\pm_x}\}$ basis and for the other $\lfloor M(K,k)\rfloor$ rounds in the basis $\{(\ket{1}+ e^{-i\pi/M(K,k)}\ket{-1})/\sqrt{2},(e^{i\pi/M(K,k)}\ket{1}- \ket{-1})/\sqrt{2}\}$ providing an estimate $\hat{\phi}_k$ of the phase $\phi_k$. After the last stage the resonator returns to a ground state of the trapping potential.}
\label{figSplittingProtocol}
\end{figure*}

We now consider the protocol outlined above to measure absolute gravitational acceleration (shown diagramatically in Figure~\ref{figSplittingProtocol}).
As depicted in Fig. \ref{figSplittingProtocol}(a), we begin with the system in the state
$\Psi(t<0) = |\alpha=0\rangle_r \, |+_x\rangle_q$,
where the subscripts $r$ and $q$ refer to the state of the resonator and qubit respectively. That is, we begin with the resonator in a harmonic oscillator ground state, and the qubit in a superposition of counter circulating currents. Then at time $t=0$ we apply the coupling Hamiltonian
$\hat{H}_{\mathrm{coupling}} = \hbar \lambda (\hat{a} + \hat{a}^{\dagger}) \hat{\sigma}^z /2$,
which imposes a constant force in the $\hat{z}-$direction on the resonator depending on the qubit state. The full Hamiltonian of the system without a  driving field on the qubit is
\begin{equation}
\hat{H} = \frac{\hbar \omega_q}{2} \hat{\sigma}^z  
           + \hbar \omega \hat{a}^{\dagger} \hat{a} + \frac{\hbar \lambda}{2} (\hat{a} + \hat{a}^{\dagger}) \hat{\sigma}^z + m g \hat{z}.
\label{eqFullCoupledH}
\end{equation}
Rewriting in the position representation using $\hat{a} + \hat{a}^{\dagger} = \hat{z}/z_0$ where $z_0=\sqrt{\hbar/2m\omega}$ gives
\begin{equation}
\hat{H} = \frac{\hbar\omega_q}{2} \hat{\sigma}^z +\frac{\hat{p}^2}{2m} + \frac{1}{2} m \omega^2 (\hat{z}-z_{\rm eq})^2
            + \frac{\hbar \lambda}{2 z_0} (\hat{z}-z_{\rm eq}) \hat{\sigma}^z + m g \hat{z},
\end{equation}
where $z = z_{\mathrm{eq}}$ is the equilibrium position of the resonator. Finally, completing the square and noting that $(\hat{\sigma}^{z})^2=1$ gives
\begin{equation}
\hat{H} = \frac{\hbar \omega_q}{2} \hat{\sigma}^z +\frac{\hat{p}^2}{2m} 
                   + \frac{1}{2} m \omega^2 \left( \hat{z} + l \hat{\sigma}^z \right)^2
                   -  m g l \hat{\sigma}^z,
\end{equation}
where we have dropped an additive constant, defined
\begin{equation}
l =\lambda z_0/\omega,
\label{eqldefn}
\end{equation}
and we have shifted the origin of the $z$ coordinate to the position $z_{\rm eq}-g/\omega^2$.
In this form, we see that because $\hat{\sigma}^z$ has eigenvalues $\pm 1$, we now have a double well potential, with the wells centred at $\pm l $ (see Figure~\ref{figSplittingProtocol}).

The resonator wave function now finds itself high on the harmonic potential slope, and experiences a state dependent force (see Fig ~\ref{figSplittingProtocol}(b)). This means the wave packet will split into a superposition of two wave packets and each of these packets will oscillate in its state dependent trap $t=\pi/\omega$ (Fig~\ref{figSplittingProtocol}(c)). 

We wait for the oscillation to complete (Fig~\ref{figSplittingProtocol}(d)), yielding the product state
\begin{equation}
\Psi(t=2\pi/\omega)  = \frac{1}{\sqrt{2}} \left( e^{i\phi} \, |1\rangle  + e^{-i\phi}|-1\rangle \right)_q\otimes |\alpha=0 \rangle_r,
\end{equation}
where the accumulated phase is 
\begin{equation}
\phi = (2 m g l -\hbar \omega_q) \, \frac{2 \pi}{\hbar \omega}.
\label{eqNBouncePhase}
\end{equation}
The expression for this phase assumes that $g$ doesn't change over the distance of oscillation. 

If we continue waiting, the system will undergo a series of $n$ oscillations and following the rapid turn off of the coupling $\lambda$ (Fig~\ref{figSplittingProtocol}(e)), we obtain the following reduced pure state for the qubit,
\begin{equation}
\hat{\rho}_q = \frac{1}{2}
   \begin{bmatrix}
    1 & e^{2in\phi} \\
    e^{-2in\phi} & 1 \\
  \end{bmatrix}.
\end{equation}
The expectation value for $\hat{\sigma}^x$ is
\begin{equation}
\langle \hat{\sigma}^x \rangle = \mathrm{Tr} \left[ \hat{\rho}_q \hat{\sigma}^x\right] = \cos(2n\phi) \equiv f(\phi)\;.
\end{equation}
When functioning as a Ramsey interferometer, the phase sensitivity we obtain by measuring the state of the qubit is given by
\begin{equation}
\Delta \phi = \frac{\delta \langle \hat{\sigma} \rangle}{\frac{d f(\phi)}{d \phi}}
                      = \frac{\sqrt{\langle \hat{\sigma}^{x2} \rangle - \langle \hat{\sigma}^x \rangle^2  } }{2 n \sin 2n \phi} 
                      = \frac{\sqrt{1-f(\phi)^2}} {2 n \sin (2n \phi)}  = \frac{1}{2n}.
\end{equation}
Referring back to (\ref{eqNBouncePhase}) we see that this means $\Delta \phi = 1/2n
                      =  4\pi m l \Delta g/\hbar \omega$,
where we have assumed  precise knowledge of $\omega,\omega_q, m, l$ (methods to pre-determine these will be detailed later). This gives an uncertainty in $g$ of
\begin{equation}
\Delta g = \frac{\hbar \omega \, \Delta \phi}{4 \pi m l}\;,
\label{eqPhaseEstimationSchemePrecisionPhi}
\end{equation}
and 
\begin{equation}
\frac{\Delta g}{g} = \frac{\hbar \omega}{8 n m g \pi l}
                                  = \frac{\hbar \omega^2}{8 \pi n\, m\,g \lambda z_0 }\;.
\end{equation}
Finally, we are constrained by the coherence time of the qubit, $\tau_c$. Specifically, we require the total evolution time to satisfy $2n\times \tau < \tau_c$, which gives
\begin{equation}
\frac{\Delta g}{g} \geq \frac{\hbar \omega}{2 \tau_c m g \lambda z_0} = \frac{\hbar}{2 \tau_c l m g}.
\label{eqdgoverglimit}
\end{equation}

\subsection{Phase estimation scheme}

The main issue with the protocol as described so far is that when measuring the phase, we only get an answer modulo $2\pi$, but the actual phase we care about is many times that, resulting a phase ambiguity. 
As an example, using the parameters in Appendix \ref{secSymbols}, we have a cat state separation of $2l=1.9$\,nm, $\omega=24.8$\,kHz and a resonator mass of $m=1.12$ nanograms, so that one slosh takes $\tau = 2\pi/\omega = 40.3\,\mu$s and the accrued phase (after subtracting the known phase $\omega_q\tau$ accumulated due to the qubit splitting) is
\begin{equation}
\phi = 2 m g l \tau / \hbar = 7.94 \times 10^{9} \, {\mathrm{rad}}\;.
\end{equation}
To solve the problem of phase ambiguity, rather than measuring $\phi$, we choose to measure a much smaller phase, arising from a much smaller displacement of the resonator. Specifically, we choose a displacement small enough such that the phase $\phi_0$ we measure during the interferometric process is $0 \leq \phi_0 < 2\pi$. We then use the nonadaptive phase estimation scheme of Ref. \cite{Higgins:2009fy} to obtain this unambiguous phase with the same degree of precision as we would if we could measure the much larger phase $\phi$ without the $2\pi$ phase ambiguity.

The scheme works by determining $\phi_0$ via successive doublings of this phase, each providing another binary bit of precision to the final estimate. Doubling is achieved by doubling the current in the resonator, resulting in twice the resonator displacement. Each of the doubled phases is measured $M$ times using the interferometric protocol described above {\color{black} but subjecting the qubit instead to a projective measurement along the $\hat{x}-$axis of the Bloch sphere at the end of each of the $M$ interferometry-measurement runs. Information from each of these measurements is used to refine the best estimate of the phase $\phi_0$ that has been obtained so far.

In detail, this works as follows: suppose we want to measure the phase $\phi_0$. We define $\phi_k = 2^k \phi_0$, with $k=0 \ldots K$. Then for each $\phi_k$ we make $M(K,k)$ interferometric measurements with specific phase offset, with $M(K,k)$ given by
\begin{equation}
M(K,k) = M_K + \mu(K-k),
\end{equation}
where $M_K$ is the number of measurements for the $2^K$ phase shift and $\mu$ is a constant. Note that because the measurement protocol is non-adaptive, the measurements can be done in any order. For each round $k$, half (or a nearest integer thereof) of the $M(K,k)$ measurements should be done in the qubit basis $\ket{\pm_x}=(\ket{1}\pm\ket{-1})/\sqrt{2}$ and the other half should be done in the $\{(\ket{1}+ e^{-i\pi/M(K,k)}\ket{-1})/\sqrt{2},(e^{i\pi/M(K,k)}\ket{1}- \ket{-1})/\sqrt{2}\}$ basis. At stage $k=K$, the phase is localised to an arc of size $2\pi/(3\times 2^K)$ and the last estimate is used as the final estimate $\hat{\phi}_0$. 

We quantify the phase uncertainty by the square root of the Holevo variance \cite{1984LNM..1055.....A}:
\begin{equation}
\begin{array}{lll}
\Delta \phi_0  \equiv  \sqrt{|\langle e^{i(\phi_0-\hat{\phi}_0)}\rangle |^{-2}-1}
\approx2\sin(|\phi_0-\hat{\phi}_0|/2)
\approx |\phi_0-\hat{\phi}_0|,
\end{array}
\end{equation}
where the approximations hold when the variance is small. It is shown \cite{Higgins:2009fy} that the precision obtained with the nonadaptive measurement protocol with the choice $M_K=2$ and $\mu=3$ provides a scaling of twice the Heisenberg limit \footnote{Specifically in Ref.~\cite{Higgins:2009fy} they report a scaling of $\Delta\phi_0<2.03 \pi/N$ for $K=20$ doublings which improves for larger $K$.}:
\begin{equation}
\Delta\phi_0{\simeq}\frac{2  \pi }{N}\;,
\label{eqDeltag1}
\end{equation}
where 
\begin{eqnarray}
N &&= \sum_{k=0}^{K} M(K,k)\, 2^k
= \sum_{k=0}^{K} (2+3(K-k))\, 2^k \nonumber \\
&&= 5 \times 2^{K+1} - 3K - 8, 
\label{eqPhaseEstimationPrecisionN}
\end{eqnarray}
which is the cumulative accrued phase in units of $\phi_0$,  during the complete estimation procedure.
To ensure the phase $\phi_0$ is less than $2\pi$ we require
$\phi_0 = 2mgl_0 \tau/\hbar < 2\pi$,
where $2l_0$ is the mean separation of the cat states, and $\tau$ is the interferometry {\it slosh} time, corresponding to one complete oscillation in the harmonic potential. Since $\tau = 2\pi/\omega$ we obtain the condition
\begin{equation}
l_0 < \frac{\hbar \omega}{2mg}.
\label{eql0defn}
\end{equation}
From Eq.~(\ref{eqldefn}) we see that
\begin{equation}
\lambda = l \sqrt{\frac{2 m \omega^3}{\hbar}},
\end{equation}
which means to get a displacement of $l=l_0$ and an associated phase $\phi=\phi_0$ we require a coupling strength of
\begin{equation}
\lambda_0 = \sqrt{\frac{\omega^5 \hbar}{2 m g^2}}.
\end{equation}
To obtain each of the doubled phases required for the protocol requires similar doubling of this coupling strength, i.e. $\lambda=2^k \, \lambda_0$ is required to produce the phase $\phi_k = 2^k \, \phi_0$.

\subsection{Parameter determination}

While our measurement protocol and phase estimation scheme gives us a phase, this phase must still be converted to a value for $g$ via Eq.~(\ref{eqNBouncePhase}). Clearly, in order to obtain a precise estimate of $g$, we must know the parameters $m, \omega,\omega_q$ and $\lambda$ to the same level of precision. These quantities can be measured offline with any additional resources, and will not affect the time taken for the phase estimation protocol.

One way to obtain information on these parameters is to observe the effect on the evolution of the qubit. If the qubit is not driven, i.e. $\Omega = 0$, then the equations of motion associated with the Hamiltonian (\ref{eqFullCoupledH}) are
\begin{equation}
\begin{split}
\frac{d\hat{a}}{dt} &= -i\omega \, \hat{a} - \frac{i \lambda}{2} \hat{\sigma}^z - i\sqrt{\frac{m g^2}{2 \hbar \omega}}  \\
\frac{{d\hat{\sigma}}_x}{dt} &= 2 \omega_q \, \hat{\sigma}^y - \lambda (\hat{a} + \hat{a}^{\dagger}) \hat{\sigma}^y \\
\frac{{d\hat{\sigma}}_y}{dt} &= -2 \omega_q \, \hat{\sigma}^x + \lambda (\hat{a} + \hat{a}^{\dagger}) \hat{\sigma}^x \\ 
\frac{{d\hat{\sigma}}_z}{dt} &= 0.
\label{eqEOM}
\end{split}
\end{equation}
If we assume the resonator starts in the ground state then we have $\langle \hat{a}(0) + \hat{a}^{\dagger}(0)\rangle = 0$. Denoting $\sigma^i = \langle \hat{\sigma}^i\rangle$ and $a = \langle \hat{a} \rangle$,  Eqs.~(\ref{eqEOM}) have the solution
\begin{equation}
\begin{split}
a(t) + a^{*}(t) &= \frac{\lambda \sigma^z (0)}{\omega} \left( \cos [ \omega t ] - 1 \right) \\
\sigma^x(t) &= \sigma^x(0) \cos \xi + \sigma^y(0) \sin \xi \\  
\sigma^y(t) &= \sigma^y(0) \cos \xi - \sigma^x(0) \sin \xi \\  
\sigma^z(t) &= \sigma^z(0)\;,
\end{split}
\end{equation}
where
\begin{equation}
\xi = 2 \omega_q t + \frac{\sigma^z(0) \lambda^2} {\omega} - \frac{\sigma^z(0) \lambda^2 \sin[\omega t]} {\omega^2}.
\end{equation}
These solutions have intricate time-dependent structure, meaning an arbitrary number of independent datapoints can be obtained by measuring, say, $\hat{\sigma}^x$ on the qubit. Provided the qubit preparation and measurement process has only statistical errors and not systematic ones, arbitrarily precise values of $\omega_q, \omega$ and $\lambda$ can be obtained by fitting a suitably large number of measurement results against the theoretically expected profile.

In order to measure the mass of the resonator, techniques such as those described by Schilling are likely to perform well \cite{Schilling:2013,Schilling:2013a}. These schemes utilize electro-optical measurement of oscillation period of a levitated superconducting oscillator, exactly the same situation as described by our scheme.

Of course, if other simpler or more precise methods are available that can provide values for any of these parameters, they can be used in the calibration process and reduce the number of parameters that need to be fitted.

\section{Precision}
\label{Sec:Precision}
Using Eq.~(\ref{eqDeltag1}) and Eq.~(\ref{eqPhaseEstimationSchemePrecisionPhi}) the gravimeter precision obtained after one full cycle of the phase estimation scheme is
\begin{equation}
\frac{\Delta g}{g} = \frac{\hbar \omega}{2 m g l_0 N},
\label{eqdgovergFunctionOfN}
\end{equation}
where $l_0$ is the cat state displacement associated with a gravitational phase $0 \leq \phi_0 < 2\pi$, and $N$, the cumulative  accrued phase over the entire cycle (in units of $\phi_0$). {\color{black} We note that through the use of this non-adaptive protocol one obtains a precision that scales as $N^{-1}$ rather than the usual $N^{-1/2}$.} Expressing $N$ in terms of the upper doubling factor $K$ using   (\ref{eqPhaseEstimationPrecisionN}) yields
\begin{equation}
\frac{\Delta g}{g} = \frac{\hbar \omega}{2 m g (l_{\rm max}/2^K)}  \frac{1}{5\times 2^{K+1}} =  \frac{\hbar \omega}{20 \, m g l_{\rm max}},
\label{eqdgOverg3}
\end{equation}
where $l_{\rm max}$ is the separation of the cat states corresponding to the maximum coupling strength $\lambda_{\mathrm{max}}=2^K\lambda_0$ generated. The optimal value of the upper doubling factor  $K$ is determined by requiring that the $K^{th}$ doubled fundamental phase $2^K\,\phi_0$ is comparable with the overall accrued phase, i.e. 
\begin{equation}
\phi_0 = \frac{1}{2^K}\frac{2 m g l_{\rm max}}{\hbar}\frac{2\pi}{\omega} < 2\pi,
\end{equation}
which means we need
\begin{equation}
K > \log_2 \left( \frac{2 m g l_{\rm max}}{\hbar \omega} \right).
\label{eqKLowerBound}
\end{equation}
This is the precision we obtain after a single phase estimation cycle incorporating $N$ projective qubit measurements. Each measurement involves initializing, evolving, and measuring the resonator over a fixed time duration which does not change throughout the cycle because we utilize a harmonic oscillator \emph{slosh} period that is constant irrespective of the spatial displacements of the wells. Thus as we execute a complete $N$-measurement estimation cycle we only alter the double well displacements via $\lambda_k$ but each of the $N$ interferometry-measurement runs take the same duration of time. This permits us to quote an effective per-root-Hertz precision if we then repeat the entire estimation cycle many times.  

In order obtain this per-root-Hertz precision, we need to know how long this phase estimation cycle takes. First consider the time $\tau_{\mathrm{exp}}$ for one interferometery run.  This consists of: (i) a qubit reset time $\tau_{\mathrm{reset}}$ to the $\ket{1}$ state, (ii) a single qubit rotation gate time $\tau_{\mathrm{rot}}$ to the $\ket{+_x}$ state, (iii) coherent evolution for one period of oscillation $\tau$, (iv) single qubit rotation $\hat{U}$ from either the $\ket{\pm_x}$ basis or the $\{(\ket{1}+ e^{-i\pi/M(K,k)}\ket{-1})/\sqrt{2},(e^{i\pi/M(K,k)}\ket{1}- \ket{-1})/\sqrt{2}\}$ basis to the $\sigma^z$ basis over a time $\tau_{\mathrm{rot}}$, and finally (v) measurement of the qubit for a time $\tau_{\mathrm{meas}}$. To obviate low frequency dephasing noise one could echo out noisy phases accumulated on off diagonal elements of the qubit state by inserting two additional steps between (iii) and (iv): (iiia) flipping the qubit state with a $\sigma^x$ gate over a time $\tau_{\mathrm{rot}}$, and (iiib) evolving the qubit for a time $\tau$ while decoupled from the oscillator, and then replacing $\hat{U}$ in step (iv) with the conjugated gate $\hat{\sigma}^x\hat{U}\hat{\sigma}^x$.  The total time for one run including the echo pulse is then 
\begin{equation}
\tau_{\mathrm{exp}}=\tau_{\mathrm{reset}}+3\tau_{\mathrm{rot}}+2\tau+\tau_{\mathrm{meas}}.
\label{timeexp}
\end{equation}
The total time required for one full phase estimation cycle is
\begin{equation}
\tau_{\phi} = \tau_{\mathrm{exp}} \sum_{k=0} ^K M(K,k) = \frac{\tau_{\mathrm{exp}}}{2} \left( 3K^2 +7K +4\right).
\end{equation}
This means the per-root-Hertz sensitivity is
\begin{eqnarray}
\frac{\Delta g}{g} \big| _{\mathrm{p.r.Hz.}} &=& \frac{\hbar \omega}{20 \, m g l_{\rm max}} \sqrt{\tau_{\phi}} \nonumber \\
&=& \frac{\hbar \omega}{20 \, m g l_{\rm max}} \sqrt{\frac{\tau_{\mathrm{exp}}}{2} \left( 3K^2 +7K +4\right)}.
\label{fullprHzeq}
\end{eqnarray}
Using (\ref{eqKLowerBound}), in the $K\gg 1$ limit we obtain 
\begin{equation}
\frac{\Delta g}{g} \big| _{\mathrm{p.r.Hz.}} \approx \frac{1}{10 \alpha} \sqrt{\frac{3 \tau_{\mathrm{exp}}} {2}} \log_2 (\alpha),
\label{eqdgOverGprHzGeneral}
\end{equation}
where 
\begin{equation}
\alpha = 2 m g l_{\rm max}/\hbar \omega.
\label{eqAlphaDefinition}
\end{equation}
So in the final analysis, the precision depends only on the parameter $\alpha$, which we want to make as large as possible. Thus we want a separation $l$ as large as possible, which in turn means engineering $\lambda$ to take values as large as possible. Lowering $\omega$ will also improve precision, but conflicts with our requirement that one slosh is completed within the coherence time of the qubit. Longer qubit coherence times would allow the precision to be improved. Finally, we want to make each preparation / evolve ({\it slosh}) / measure sequence as quick as possible.

To obtain a quantative estimate of the precision our scheme can achieve, we assume the system parameters listed in Table \ref{default}. Using (\ref{eql0defn}) we see that with our assumed parameters we require $l_0 < 0.75\times 10^{-18}$\,m. 
This means we have
$K = \log_2 \left( l_{\rm max}/l_0 \right) = 30.2$,
indicating that we perform 31 doublings during the phase estimation protocol, increasing our initial phase from $\phi_0$ to $2^K \phi_0$.

We take thickness of the resonator loop wire as a free parameter to be chosen in fabrication. Changing this parameter has the effect of changing the mass of the resonator, which in turn alters the mechanical oscillation frequency. The frequency as a function of wire radius for our assumed system parameters is plotted as the red curve in Figure~\ref{fig_omega_and_precision_vs_wire_size}. Changing the resonator mass will also change the precision, as it affects all the parameters $\omega, \lambda, z_0$, and $l$. The per-root-Hertz precision as a function of the resonator wire thickness is plotted as the blue curve in Figure \ref{fig_omega_and_precision_vs_wire_size}.
\begin{figure}[htb]
\centering
\includegraphics[width=8cm]{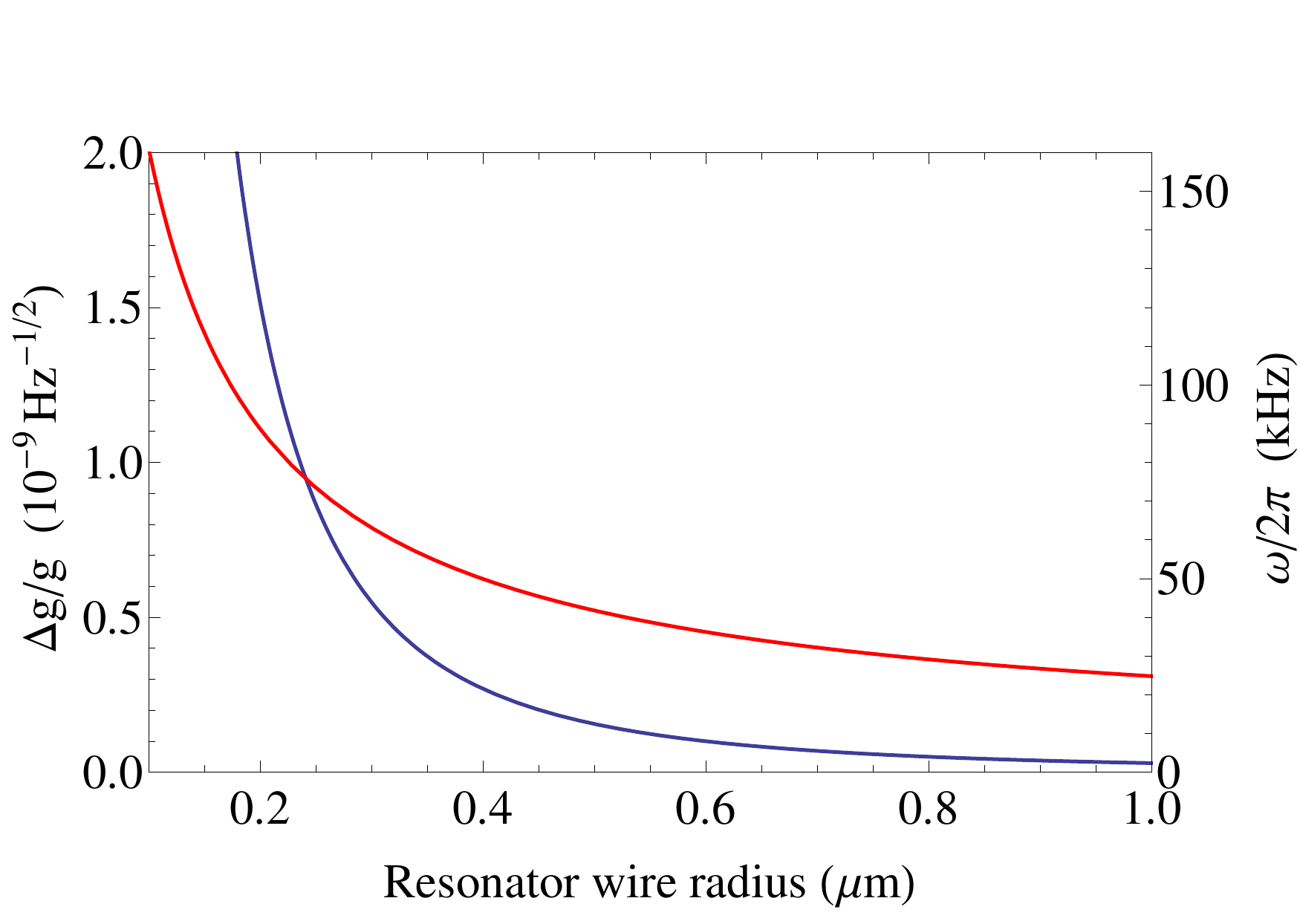}
\caption{Mechanical oscillation frequency $\omega$ (red) and the per-root-Hertz sensitivity of our gravimeter in the absence of decoherence (blue) as a function of the resonator loop wire radius. Increasing the radius increases the mass of the resonator, which in turn reduces the oscillation frequency. Sensitivity increases rapidly as the thickness increases. This is due to a larger radius wire giving the resonator a larger mass and increasing the oscillation period. This in turn results in a strong coupling to the gravitional field, and a longer time spent sampling that field. System parameters are as given in Table \ref{default}.}
\label{fig_omega_and_precision_vs_wire_size}
\end{figure}
This precision compares favourably with the best free-fall corner cube measurements ($\Delta g/g = 1.5\times 10^{-8}\,{\mathrm{Hz}}^{-1/2}$ \cite{1995Metro..32..159N}) and cold atom interferometers ($\Delta g/g = 4.2\times 10^{-9}\,{\mathrm{Hz}}^{-1/2}$ \cite{Hu:2013kq}). We note, however, that this precision was obtained in the limit of no decoherence; we will consider the effect of decoherence on our sensitivity in Section \ref{secDecoherenceEffects}.



\section{Decoherence} 
\label{secDecoherence}

We now review the various potential sources of decoherence and their effects on the performance of the gravimetry protocol.

\subsection{Quality Factor}
In the subsequent discussions we make  use of the quality factor $Q$ of the mechanical oscillations of our resonator. One usual definition of $Q$ is given by
\begin{equation}
Q= \frac{\hbar \omega^2}{P},
\label{eqQdefgroundstate}
\end{equation}
where $P$ is the power loss, $\omega$ is the oscillation frequency and $\hbar \omega$ is the energy of the system. In our protocol, however, the resonator is not in the motional ground state --- the resonator is oscillating back and forth with a large amplitude (several nanometers). As we begin the interferometry protocol (or {\it slosh}), with the resonator high up on a potential hill, we have $V(l) = \frac{1}{2} m \omega^2 l^2$
where $l=\lambda z_0 / \omega$ is the displacement from equilibrium, and $z_0 = \sqrt{\hbar/2 m \omega}$ is the harmonic oscillator ground state extent. This means for our system we have
$V = \hbar \lambda^2/4\omega$.
Associating this potential energy with the energy in (\ref{eqQdefgroundstate}) we obtain
\begin{equation}
Q \approx \frac{\hbar \lambda^2}{4 P}.
\label{eqOurQ}
\end{equation}
We also note that for all the calculations in this section we use the system parameters described in Appendix \ref{secSymbols}.

\subsection{Qubit dephasing}
 The effect of qubit decoherence on the evolution of the joint qubit-resonator system is solved for in Appendix \ref{Decoherence}. The main result is that the off diagonal elements of the qubit density matrix, which carry the gravitationally induced phase accumulation, will experience exponential decay by a factor $e^{-\tau/T_2}$ where $T_2$ is the qubit dephasing time. Dephasing rates vary greatly with the superconducting circuit architecture. Recent experiments with superconducting flux qubits in 3D microwave cavities have reported decoherence times of $T_{2}^{\rm echo}>19\,\mu$s \cite{Stern:2014fk}, while decoherence times of $T_{2}^{\rm echo}>100\,\mu$s have been reported for transmon qubits in 3D cavities \cite{Geerlings:2013}.

\subsection{Decoherence due to eddy currents in the magnet}

As the resonator oscillates, carries currents, and is in close proximity to the magnetised sphere, it will inductively induce eddy currents in the sphere, which will result in power loss as the magnetic material has electrical resistance. In order to estimate this effect we consider infinitesimal horizontal loops of radius $R'$ inside the sphere
and placed at a distance $h$ from the bottom of the sphere. The electromotive force induced in each of
such loops due to the resonator motion is given by $|\epsilon| = M_{l,s}(R',h) \, dI_r/dt$, where $M_{l,s}$ is the mutual inductance between the horizontal loop of the resonator and the horizontal infinitesimal loop of the sphere.This gives an upper bound on the power loss as
\begin{eqnarray}
P &\leq& \int \frac{\epsilon^2}{\rho 2 \pi} dR' \, dh \nonumber \\
    &=& \left( \frac{\mu_0}{4\pi} \right)^2 \frac{2 \pi^3 R_r^4 I_r^2 \omega^2}{\rho }\int_{h=0} ^{2R_s} dh \int _0 ^{\sqrt{R_s^2 -(R_s-h)^2}} dR' \frac{R'^3}{(r_0 + h)^6} \nonumber \\ 
&=& \left( \frac{\mu_0}{4\pi} \right)^2 \frac{2 \pi^3 R_r^4 I_r^2 \omega^2}{\rho } \frac{4 R_s^5}{15 r_0 ^3 (r_0+2R_s)^3},
\end{eqnarray}
where $r_0$ is the minimum distance from the bottom of the sphere to the centre of the resonator, and $\rho$ is the resistivity of the magnetic material. Our sphere is composed of YIG, which has $\rho=10^{12}\,\Omega$m; we take the $I_r$ to be the largest current reached in the resonator (occurring at full displacement), i.e. $I_{r{\mathrm{max}}}{\sim}\, 48\,\mu$A.
This gives the power loss due to eddy currents in the YIG sphere as $P=6.2 \times 10^{-38}$\,W, which via (\ref{eqOurQ}) corresponds to a quality factor of $Q=\hbar \omega^2/P$ = $3.1 \times 10^{22}$.

\subsection{Dipole radiation}

An oscillating loop carrying current will emit electromagnetic radiation, dissipating energy from our system. We treat our resonator loop as a dipole, with a current given by $I=I_{r{\mathrm{max}}} e^{i \omega t}$. The power loss of an oscillating dipole due to radiation is given by
$P =  R_{\mathrm{rad}} I^2/2$,
where
$R_{\mathrm{rad}} = \frac{\pi}{6} \left( \frac{R_r \omega}{c} \right)^4 Z$,
where $Z=377 \, \Omega$ is the impedance of the vacuum. Using the parameters in Appendix \ref{secSymbols} we obtain a power loss of $P=2.5 \times 10^{-41}\,$W, corresponding to $Q=7.5 \times 10^{25}$.

\subsection{Background gas collisions}
\label{GasCollisions}
In the limit where the mean free path of the gas molecules is sufficiently large, the damping rate is given by \cite{Guccione:2013fv}
\begin{equation}
\Gamma = 2\, \rho_{\mathrm{gas}}\, A\, u_{\mathrm{av}} / m_g,
\label{eqGasLambda}
\end{equation}
where $\rho_{\mathrm{gas}}$ is the density of the gas, $A= 2\pi R_r 2 a $ is the cross-sectional area of the resonator interacting with the gas, $m_g$ is the mass of a gas molecule, and $u_{\mathrm{av}} = \sqrt{2 k_B T / m_g}$ is the average velocity of a gas molecule. In order to be in this limit, the system must have a Knudsen number ${\mathrm{Kn}} > 10$ \cite{Bhiladvala:2004jt}. Using the parameters in Appendix \ref{secSymbols}, our system has ${\mathrm{Kn}}\,{\sim}\, 10^{9}$, far into regime where Eq.~(\ref{eqGasLambda}) is valid.
Taking $m_g=3.98 \times 10^{-26}$\,kg (nitrogen molecule), $T=0.1$\,K, a resonator ring radius of $5\,\mu$m, a resonator wire radius of $1.0\,\mu$m, and area of $A=6.28\times 10^{-11}$\,m$^2$, and a pressure of $10^{-9}$\,Pa, Eq.~(\ref{eqGasLambda}) yields $\Gamma_{\mathrm{gas}}=2.7 \times 10^{-8}$\, Hz and an associated $Q=9.2 \times 10^{11}$.

\subsection{Coupling to torsional modes}

Coupling of the vertical ($z$) centre of mass ($z-$COM) oscillation mode to other bending/twisting/torsional modes of the resonator also allows energy to leak from the $z-$COM phonon mode. Of these alternative motional modes one can  consider, the lowest frequency mode is the torsional mode which has frequencies
\begin{equation}
\nu = \frac{1}{2 \pi} \sqrt{\frac{E A}{2 \mu R_r^2}} \sqrt{1+n^2} \; ,
\end{equation}
where $n > 0$ is the integer valued mode number, $A$ is the cross sectional area of the wire, $\mu=\rho \pi a^2$ is the mass per unit circumference, and $E$ is the Young's modulus of the wire. We take $E=16 \times 10^9$\,Pa, $\rho=11340$\,kg/m$^3$ giving $\mu = 3.56 \times 10^{-8} $\,kg/m. This gives the lowest frequency mode as $\nu = 1.89 \times 10^{8}$\,rad/s, which is ${\sim}\, 1200$ times larger than $\omega$, indicating cross-coupling to other modes is negligible.

\subsection{Effects of decoherence on the protocol}
\label{secDecoherenceEffects}

We now examine  the effects of decoherence on the joint state of the resonator and the qubit as well as the effects of noise during the qubit preparation and readout stages. We assume a motional damping environment for the resonator and a damping and dephasing environment for the qubit. These  are the dominant sources of decoherence in the system.

Other error processes include: noisy qubit initialisation in state $\ket{1}_Q$, noisy implementation of a qubit unitary rotation $\hat{U}$, and imperfect measurement. 
Noisy initialization can be modelled as erroneously preparing a mixed input state by mixing in the complement to the ideal state with probability $p_{\rm init}$ described by the map: $\mathcal{E}^{({\rm init})}=(1-p_{\rm init})\ket{1}\bra{1}+p_{\rm init}\ket{-1}\bra{-1}$.
A noisy qubit rotation is modelled as a map where with probability $1-p_{\rm rot}$ the correct unitary opaerator $\hat{U}$ is applied and with probabiliy $p_{\rm rot}$ the qubit is completely depolarized:  $\mathcal{E}^{({\rm rot})}(\hat{\rho})=(1-p_{\rm rot})\hat{U}\hat{\rho}\hat{U}^{\dagger}+p_{\rm rot}{\bf 1}_2$,
Noisy measurement is modelled as flipping the qubit with some probability $p_{\rm meas}$ before performing a perfect measurement:  $\mathcal{E}^{({\rm meas})}(\hat{\rho})=(1-p_{\rm meas})\hat{\rho}+p_{\rm meas}\hat{\sigma}^x \hat{\rho}\hat{\sigma}^x$.  In a spin echo sequence, the qubit would be coupled to the resonator for a time $\tau=2\pi/\omega$ described by the map $\mathcal{E}^{({\rm ev A})}$, then the coupling would be set to zero, the qubit would be flipped with a $\hat{\sigma}^x$ gate, and the system would freely evolve for a period $\tau$ described by the map $\mathcal{E}^{({\rm ev B})}$.   The composition of all these error processes in a full spin echo sequence gives a final output measurement of the desired value of $\cos(\phi)$ of
\begin{equation}
\begin{array}{lll}
\langle \hat{\sigma}^x\rangle&=&\tr[\hat{\sigma}^x \mathcal{E}^{({\rm meas})}\circ \mathcal{E}^{({\rm rot})}\circ \mathcal{E}^{({\rm ev B})}\circ\mathcal{E}^{({\rm rot})}\\
&&\circ \mathcal{E}^{({\rm ev A})}\circ \mathcal{E}^{({\rm rot})}\circ \mathcal{E}^{({\rm init})}(\rho)]  \nonumber \\
&=& f\cos(\phi).
\label{46}
\end{array}
\end{equation}
where we have introduced the cumulative per round fidelity
\begin{equation}
f=e^{-4\pi/\omega T_2}e^{-4\pi\Gamma l^2/z_0^2\omega}(1-2 p_{\rm init})(1-p_{\rm rot})^3(1-2 p_{\rm meas}).
\label{eqFidelityDefinition}
\end{equation}
This form for the fidelity is valid when the rotation gate times and measurement times are small compared to the period of 
the resonators oscillation which is usually the case. If not then the factor $e^{-4\pi/\omega T_2}$ should be replaced by
$e^{-\tau_{\rm exp}/T_2}$.

To determine the decoherence rate $\Gamma$ we sum the rates $\Gamma_i$ for resonator damping described above using $\Gamma_i = \omega/2 \pi Q_i$, with the quality factors $\{Q_i\}$ factors taken from the previous sections. This gives us $\Gamma_{\mathrm{rad}}= 3.3 \times 10^{-22}$\,s$^{-1}$ (dipole radiation); $\Gamma_{\mathrm{eddy}} = 8.1 \times 10^{-19}$\,s$^{-1}$ (eddy currents in sphere); and $\Gamma _{\mathrm{gas}} = 2.7 \times 10^{-8}$\,s$^{-1}$ (background gas collisions).
Putting these damping rates into Eq.~(\ref{eqDecoherenceFactor}) yields the following decoherence factors after a single resonator oscillation with the maximum separation:
\[
\frac{4\pi l_{\rm max}^2}{z_0^2\omega}\times [\Gamma_{\mathrm{rad}},\;\Gamma_{\mathrm{eddy}},\; \Gamma_{\mathrm{gas}}]{\sim}\, [ 7.9 \times 10^{-17}, \; 1.9 \times 10^{-13},\; 6.5 \times 10^{-3}], \]
indicating that collisions with background gas molecules is the most significant form of amplitude damping. However even then this damping yields a 99.4\% fidelity after a single oscillation period.

Each stage $k$ of our protocol involves estimating the value of the phase by estimating the probability the qubit is in state $\ket{M}$, i.e. an estimation of $p_M=(1\pm\cos(\phi_k))/2$. Given the reduced polarisation of the qubit due to errors (Eq. \ref{46}), the procedure is akin to estimating the probability $p$ that a biased coin lands heads subject to noise such that each observation gets flipped with probability $p_{\rm noise}=(1-f)/2$. This scenario of estimating the bias of a noisy coin was studied in Ref. \cite{Ferrie:2012} where it was shown that a hedged maximum likelihood method provides a good estimate of an unknown $p$ given a known $p_{\rm noise}$. The effect of the reduced visibility due to finite fidelity is to increase the number of measurements per stage, $M(k,K)$ by a factor of $1/f^2$ in order to keep the same overall precision of our protocol.  Note this multiplicative factor is independent of the stage $k$ since the the operation time always involves single sloshes whose period is solely determined by the resonator frequency $\omega$. The overall effect on the precision is then
\begin{equation}
\frac{\Delta g}{g} \big| _{\mathrm{p.r.Hz.}} =\frac{\hbar \omega}{10 f \, m g l_{\rm max}} \sqrt{\frac{\tau_{\mathrm{exp}}}{2} \left( 3K^2 +7K +4\right)}\;.
\label{fullprHzeqwitherror}
\end{equation}
In order to determine the fidelity, we need to know qubit operation times, qubit error rates, and dephasing time. Recent experiments using superconducting transmon qubits in three dimensional microwave have shown dephasing times of $T_2^{\mathrm{echo}} = 70\,\mu$s, reset times $\tau_{\mathrm{reset}}= 3\,\mu$s and error rates $p_{\rm reset}\leq 0.005$ \cite{Geerlings:2013}. All the other operations needed for fault tolerant quantum computation have been demonstrated with superconducting qubits as well. In Ref. \cite{Chow:1en} the following operation times and errors were reported for transmon qubits:  $\tau_{\mathrm{rot}}= 40$\,ns, $\tau_{\mathrm{meas}}=4\,\mu$s, $p_{\rm rot}\leq 0.003$, $p_{\rm meas}\leq 0.09$.

%


Assuming a flux qubit with the same operation times and using Eq.~(\ref{eqFidelityDefinition}) along with Eq.~(\ref{timeexp}) to obtain $\tau_{\mathrm{exp}}$, Eq.~(\ref{fullprHzeqwitherror}) allows us to determine the ultimate sensitivity of our gravimeter, taking into account qubit errors, readout and preparation time and decoherence. The result is plotted in Figure~\ref{fig_precision_vs_wire_size} showing for a resonator wire thickness of $1\mu$m an achievable precision of $\frac{\Delta g}{g} \big| _{\mathrm{p.r.Hz.}}= 2.21\times 10^{-10}$\,Hz$^{-1/2}$. Even with the decrease in fidelity as we increase resonator mass, the per-root-Hertz precision still increases monotonically with this increase, albeit at a slower rate than the perfect decoherence-free case shown in Figure~\ref{fig_omega_and_precision_vs_wire_size}. 

\begin{figure}
\centering
\includegraphics[width=8.5cm]{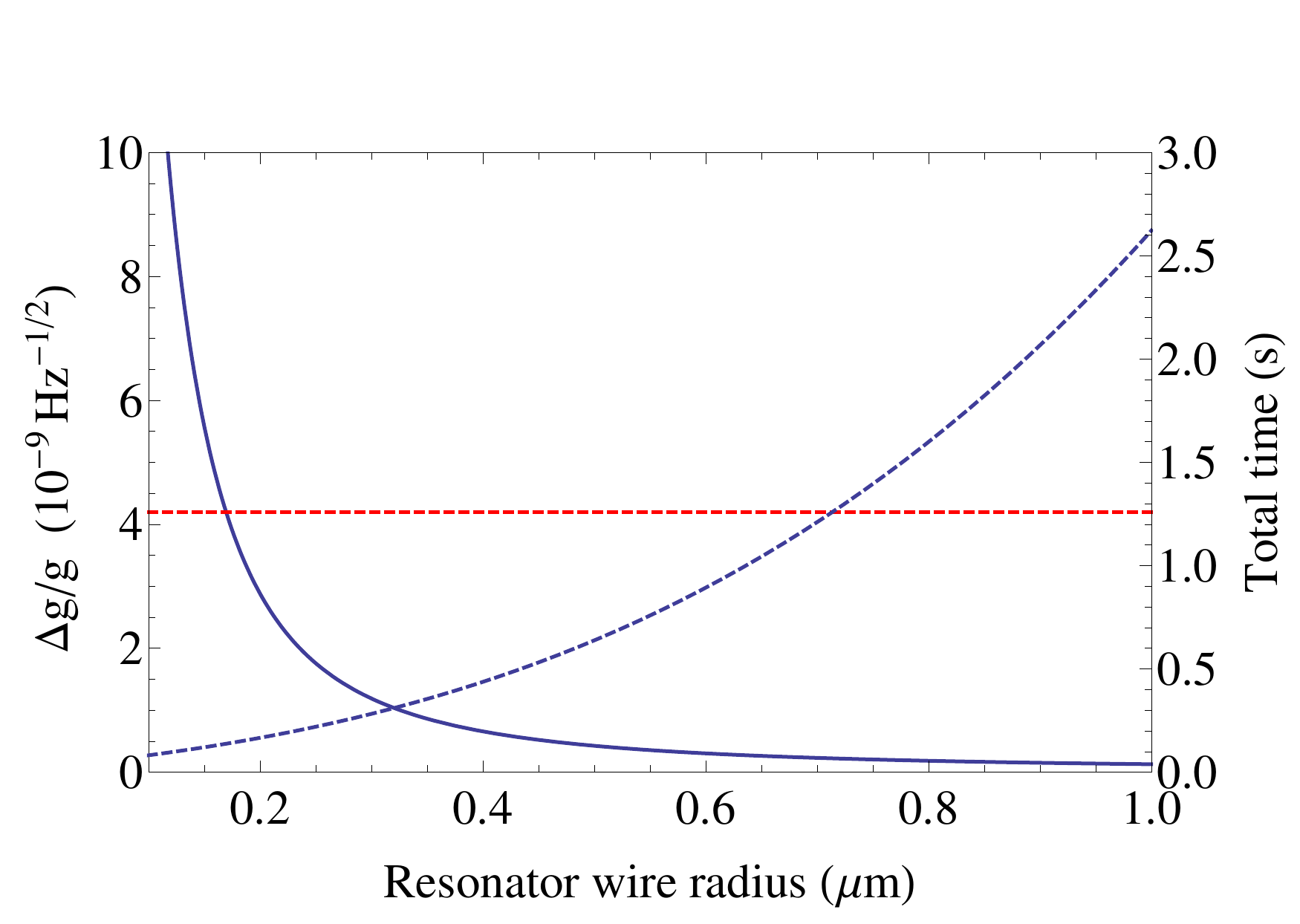}
\caption{Performance of our gravimeter as a function of the resonator loop wire radius taking into account current experimental preparation, readout, and dephasing times. Solid blue line shows per-root-Hertz precision; dashed blue line shows the time required for a full $\Delta g/g$ measurement, taking into account the requirement for more measurements as fidelity decreases; red dashed line indicates current best absolute gravimeter precision \cite{Hu:2013kq}. In principle precision can be increased without limit, but at some point long term equipment drift or the timescale of the phenomenon of interest will become an issue. For that reason we take a $1\,\mu$m resonator wire radius as a plausible upper limit. System parameters are as given in Table \ref{default}. }
\label{fig_precision_vs_wire_size}
\end{figure}

\section{Conclusion}
\label{Gravconc}

We have presented a scheme for absolute gravimetry utilising quantum magnetomechanics and Schr{\"{o}}dinger cat states. The protocol interferometrically measures the differential gravitational phase accrued between the two heights of a macroscopic quantum resonator placed into a vertical spatial superposition. With realistic materials and current reported values for superconducting qubit coherence times we obtain a sensitivity of $\Delta g/g = 2.21\times 10^{-10}$\,Hz$^{-1/2}$ for the thickest resonator wire we considered, which is over an order of magnitude better than the $\Delta g/g = 4.2\times 10^{-9}$\,Hz$^{-1/2}$ achieved by current state-of-the-art absolute gravimeters which rely on atom interferometry \cite{Hu:2013kq}. Furthermore, this sensitivity can be substantially improved on, primarily by improving the coherence time of the flux qubit, but also by using lower temperatures and more complicated magnet-resonator geometries.

Our scheme involves the production a series of Schr{\"{o}}dinger cat states, the largest of which is a superposition of $1.1 \times 10^{-12}$\,kg masses displaced by ${\sim}\, 10^5$ times the width of their center of mass wave function. While these are very fat cats by Schr{\"{o}}dinger cat standards, the result is that the gravity measurement is made over a distance of only $1.9\times 10^{-9}$\,m, allowing the technique to probe spacial regions eight orders of magnitude smaller than current schemes involving springs, falling corner cubes and atom interferometry  \cite{Crossley:2013bj}.

The precision is constrained by the dynamic range of the qubit-resonator coupling parameter, as well as the coherence time of the qubit. The coupling strength is limited at the low end by the current noise floor of the qubit, and at the high end by the critical current value of the qubit and the inhomogeneity of the magnetic field of the sphere levitating the resonator. We chose a spherical geometry for the magnet as this allowed analytic results, but there is certainly scope to generate fields with higher inhomogeneities through more complicated geometries. The noise floor of the qubit is largely governed by its temperature; we have assumed a temperature of $0.1$\,K. Lower temperatures would be challenging but would proportionally increase sensitivity.

It is likely that improving the dephasing coherence time of the qubit is the best route to improved sensitivity, as this is the dominant source of decoherence. Longer coherence times allow for longer oscillation periods of the resonator which can easily be arranged by increasing its mass, allowing both a larger coupling to the gravitation field and a longer time spent sampling that field over a single oscillation.

\section*{Acknowledgements}
We would like to acknowledge helpful discussions with John Clark and Ray Simmonds. This work was partly supported by the ARC Centre of Excellence for Engineered Quantum Systems EQUS (Grant No CE110001013).


\newpage
\appendix
\newpage
\section{Symbols and values}
\label{secSymbols}
\begin{table*}[htdp]
\caption{System parameters and the values used in the main text for precision gravimetry.}
\begin{center}
\begin{tabular}{lclll}
\hline
Symbol & & Value & Definition \\
\hline
$\Phi_0$          &=& $2.07 \times 10^{-15}$\,Wb     & flux quantum \\
$g$               &=& 9.81\,m\,s$^{-2}$              & acceleration due to gravity \\
$m$               &=& $1.12\times 10^{-12}$\,kg      & resonator mass (Pb) \\
$\omega/2\pi$     &=& $24.8$\,kHz                    & resonator frequency \\
$z_0$             &=& $1.74\times 10^{-14}$\,m       & ground state rms width of resonator \\
$R_q$             &=& $5\,\mu$m                      & radius of qubit loop \\
$R_r$             &=& $5\,\mu$m                      & radius of resonator ring \\
$R_{\rm sphere}$  &=& $10\,\mu$m                     & radius of magnetized sphere\\
$a$               &=& $1.0\,\mu$m                    & radius of resonator wire \\
$d$               &=& $2.0\,\mu$m                    & distance between resonator centre of mass and qubit\\
$r_0$             &=& $1\,\mu$m                      & minimum distance from sphere surface to centre of mass of resonator \\
$z_{\mathrm{eq}}$ &=& $11\,\mu$m                     & equilibrium position of resonator \\
$V$               &=& $4.19 \times 10^{-15}$\,m$^3$  & volume of magnetised sphere \\
$\mathcal{M}$     &=& $8.76\times 10^2$\,A\,m$^{-1}$ & magnetisation of YIG sphere \\
$\rho$            &=& $10^{12}\,\Omega$m             & resistivity of magnetised sphere made of YIG \\
$l_{\rm max}$     &=& $9.5\times 10^{-10}$\,m       & largest size of Schr\"odinger cat \\
$\lambda_{\rm max}/2\pi$ &=& $1.35$\,GHz             & maximum qubit-resonator coupling\\
$\lambda_{0}/2\pi$&=& $0.63$\,Hz                      & minimum qubit-resonator coupling\\
$L_r$             &=& $2.25 \times 10^{-11}$\,H      &  resonator self inductance\\
$L_{q}$           &=& $1.38 \times 10^{-11}$\,H      & qubit self inductance \\
$M_{rq}$          &=& $6.75 \times 10^{-12}$\,H      & mutual inductance between resonator and qubit  \\
$\omega_q/2\pi$   &=& $6$\,GHz                     & qubit energy level splitting \\ 
$\Phi$            &=& $2.37 \times 10^{-12}$\,Wb     & flux through the resonator  \\
$T_q$               &=& $100$\,mK                    & temperature of qubit system \\ 
$I_{q {\rm max}}$ &=& $75\,\mu$A                     & maximum current in qubit \\
$I_{q 0}$         &=& $3.5\times 10^{-14}\,$\,A      & minimum current in qubit \\
$I_{r {\rm max}}$ &=& $48\,\mu$A                     & maximum current in resonator\\
$\tau_{\mathrm{exp}}$ &=& $87.8\,\mu$s               & time for one complete prepare / evolve / measure run \\
$\tau_c$          &=& $70 \,\mu$s                    & coherence time of the qubit \\
$T_1$              &=& $70\,\mu$s                      & qubit $T_1$ coherence time \\
$T_2$              &=& $70\,\mu$s                      & qubit $T_2$ coherence time \\
$\Gamma_{\rm gas}/2\pi$ &=& $2.7\times 10^{-8}$\,Hz  & resonator amplitude damping rate due to background gas collisions \\ 
$\Gamma_{\rm eddy}/2\pi$ &=& $8.1 \times 10^{-19}$\,Hz & resonator amplitude damping rate due to induced eddy current losses \\ 
$\Gamma_{\rm rad}/2\pi$ &=& $3.3\times 10^{-22}$\,Hz & resonator amplitude damping rate due to magnetic dipole radiation \\ 
 \hline 
 \end{tabular}
\end{center}
\label{default}
\end{table*}
\newpage

\section{Open System Dynamics}
\label{Decoherence}

The open systems dynamics of the joint qubit-resonator system is given by the master equation
\begin{equation}
\dot{\hat{\rho}}(t)=\hat{\mathcal{L}}(\rho(t))
\end{equation}
with the Louivillian
\[
\hat{\mathcal{L}}=-\frac{i}{\hbar}[\hat{H},\rho(t)]+\hat{\mathcal{L}}_r+\hat{\mathcal{L}}_q.
\]
Free evolution is governed by the Hamiltonian
\[
\hat{H} = \frac{\hbar \omega_q}{2} \hat{\sigma}_z +\frac{\hat{p}^2}{2m} 
                   + \frac{1}{2} m \omega^2 \left( \hat{z} + l \hat{\sigma}_z \right)^2
                   -  m g l \hat{\sigma}_z,
\]
and amplitude damping of the resonator and amplitude and phase damping of the qubit are described by:
\begin{equation}
\begin{split}
\hat{\mathcal{L}}_r&= \frac{\Gamma}{2}\hat{D}[\hat{a}] \\
\hat{\mathcal{L}}_q&= \frac{\Gamma_{\perp}}{2} (N_q+1) \hat{D}[\hat{\sigma}^-]+ \frac{\Gamma_{\perp}}{2} N_q \hat{D}[\hat{\sigma}^+]+\frac{\Gamma_{\parallel}}{4} \hat{D}[\hat{\sigma}^z]\;,
\end{split}
\end{equation}
with the map $\hat{D}$ defined as 
\[
\hat{D}[\hat{O}](\hat{\rho})\equiv 2 \hat{O}\hat{\rho} \hat{O}^{\dagger}-\{ \hat{O}^{\dagger} \hat{O},\hat{\rho}\}.
\]
The equilibrium phonon occupation of the qubit environment is $N_q=(e^{-\hbar \omega_q/k_B T_q}-1)^{-1}$ where $T_q$ is the qubit phonon bath temperature. The decay rates are related to the usual decoherence times according to 
$T_1^{-1}\equiv \Gamma_{\perp}(2N_q+1)$ and $T_2^{-1}\equiv T_1^{-1}/2+\Gamma_{\parallel}$.
We treat the environment of the resonator as zero temperature meaning the resonator only loses energy to the environment. This is justified as it is not clamped to any material and we assume the surrounding cavity is in the electromagnetic vacuum state.  Any temperature dependence of damping due to background gas collisions can be encorpeated into the value of damping rate $\Gamma_{\rm gas}$ as described in Sec. \ref{GasCollisions}. 

At each measurement run, the joint state of the qubit and resonator is prepared in the initial state
\begin{equation}
\hat{\rho} (0)=\frac{1}{2} \left(\begin{array}{cc}  1 & 1 \\ 1 & 1 \end{array}\right)_q \otimes \ket{0}_r\bra{0}\;,
\end{equation}
where $\ket{0}_r$ is the motional ground state of the resonator.  
At this point we can make some simplifications. We are interested in obtaining a worst case scaling for the decoherence of our protocol which would occur when the size of the initial Schr\"odinger cat state is largest, i.e. $l=l_{\rm max}$. The time evolution is only over one period of oscillation $\tau=2\pi/\omega$ of the resonator and we assume that $\Gamma_{\perp},\Gamma_{\parallel}< \omega$ and $\Gamma\ll \omega$. It is convenient to divide the Louivillian into two parts: $\hat{\mathcal{L}}=\hat{\mathcal{L}}_1+\hat{\mathcal{L}}_2$:
\[
\hat{\mathcal{L}}_1=-\frac{i}{\hbar}[\hat{H},\cdot]+\hat{\mathcal{L}}_r+\frac{\Gamma_{\parallel}}{4} D[\hat{\sigma}^z]
\]
and
\[
\hat{\mathcal{L}}_2=\frac{\Gamma_{\perp}}{2} (N_q+1) D[\hat{\sigma}^-]+ \frac{\Gamma_{\perp}}{2} N_q D[\hat{\sigma}^+]
\]
During evolution by $\hat{\mathcal{L}}_1$, the operator $\hat{\sigma}^z$ is a conserved quantity and we can solve for the joint evolution of the qubit and resonator exactly.  Evolution by $\hat{\mathcal{L}}_2$ describes pure amplitude damping of the qubit.  We then approximate the evolution of the system over one resonator oscillation period $\tau$ as the composition of maps:
\[
\mathcal{E}^{({\rm evA})}(\hat{\rho}(0))\equiv e^{\hat{\mathcal{L}}\tau}(\hat{\rho}(0))\approx e^{\hat{\mathcal{L}}_2 \tau}\circ  e^{\hat{\mathcal{L}}_1 \tau}(\hat{\rho}(0)).
\]

We first consider evolution by $\hat{\mathcal{L}}_1$. The qubit dephasing simply introduces decay of off diagonal qubit states. Damping maps coherent states to coherent states and since we begin in a superposition of coherent states, at any time $t$ we can write the joint state in the interaction picture $\hat{\rho}_I(t)=e^{i\hat{H}t}\hat{\rho}e^{-i\hat{H}t}$ as
\[
\hat{\rho}_I(t)=\sum_{M,M'=-1}^1 c_{M,M'}e^{-\frac{\gamma_Q}{2} |M-M'|t}\ket{M}\bra{M'}\otimes \hat{A}^{M,M^{\prime }}_I(t),
\]
where the eigenbasis of $\hat{\sigma}^z$ is $\ket{M=\pm 1}$ and 
\[
\hat{A}^{M,M^{\prime }}_I(t)= \ket{\alpha^M_{I}(t) }\bra{\beta^{M'}_{I}(t)}\;.
\]
To derive the evolution during decay we use the characteristic function
\[
X(t)=\mbox{Tr}_R [A^{M,M^{\prime }}_I(t)e^{\Lambda
\hat{a}_I^{\dagger}}e^{-\Lambda^{\ast}\hat{a}_I}]\;,
\]
where the trace is taken over the resonator's motional degree of freedom such that
\begin{align*}
\dot{X}(t) & = \mbox{Tr}_R[\dot{\hat{A}}^{M,M^{\prime }}_I(t) e^{\Lambda
\hat{a}_I^{\dagger}}e^{-\Lambda^{\ast}\hat{a}_I}] \\
& = \Gamma \mbox{Tr}_F [
(\hat{a}_I\hat{A}^{M,M^{\prime }}_I(t)\hat{a}_I^{\dagger} \\
& \phantom{=} -\frac{1}{2}\hat{a}_I^{\dagger}\hat{a}_I \hat{A}^{M,M^{\prime }}_I(t) -\frac{1}{2}
\hat{A}^{M,M^{\prime }}_I(t)\hat{a}_I^{\dagger}\hat{a}_I) e^{\Lambda
\hat{a}_I^{\dagger}}e^{-\Lambda^{\ast} \hat{a}_I}]\;.%
\end{align*}
Using the relations
\[
e^{-\Lambda^{\ast}\hat{a}}\hat{a}^{\dagger}=(\hat{a}^{\dagger}-\Lambda^{\ast})e^{-\Lambda^{%
\ast}\hat{a}},\quad \hat{a}e^{\Lambda \hat{a}^{\dagger}}=e^{\Lambda \hat{a}^{\dagger}}(\hat{a}+\Lambda)\;,
\]
we obtain
\begin{align}
\dot{X} & = -\frac{\Gamma}{2}\Big(\Lambda^{\ast}%
\frac{\partial \hat{X}}{\partial \Lambda^{\ast}}+\Lambda \frac{\partial \hat{X}}{%
\partial \Lambda}\Big) \nonumber\\
& = -\frac{\Gamma}{2}\Big(\beta^{M'\ast}_{I}(t)\Lambda-%
\alpha^{M}_{I}(t)\Lambda^{\ast}\Big)X\;.%
\label{Xd}
\end{align}

To solve for the dynamics, we make the ansatz:
\begin{equation}
X(t)=Ce^{-\lambda ^{\ast }\alpha^M_I(t)}e^{\lambda \beta^{M'\ast}_I(t)}\;.
\label{anz}
\end{equation}%
From the reflection symmetry of the state dependent traps, the magnitudes of the coherent states correlated with the qubit states are equal at all time so 
we can write $\beta^{M'}_I(t)=\alpha^{M'}_I(t)$.  Evaluating the time derivative of $X(t)$ and setting this equal to Eq. \eqref{Xd} we have 
\begin{align*}
\alpha^M_I(t) & = (e^{-\Gamma t/2}+1)\frac{\lambda M}{2\omega}\;.\\
\end{align*}%
This solution simply reflects the fact that the initial coherent state for the spatially localised oscillator with mean position $gz_0/\omega^2$ is a displaced coherent state with respect to the potential minimum with respect to qubit state $M$ of the Hamiltonian $\hat{H}$.  Using the characteristic equation we can write 
\[
X(t)=\mbox{Tr}_R [A^{M,M^{\prime }}_I(t)e^{\Lambda
\hat{a}_I^{\dagger}}e^{-\Lambda^{\ast}\hat{a}_I}]=\mbox{Tr}_R [A^{M,M^{\prime }}_I(0)e^{\Lambda
\hat{a}_I^{\dagger}(t)}e^{-\Lambda^{\ast}\hat{a}_I(t)}]\;.
\]
The diagonal terms evolve as 
\[
e^{\hat{\mathcal{L}}_r t}[\ket{\alpha^{M}_I(0)}\bra{\alpha^{M}_I(0)}]=\ket{\alpha^{M}_I(t)}\bra{\alpha^M_I(t)}\;.
\]
The off-diagonal terms evolve as
\begin{equation}
\begin{split}
e^{\hat{\mathcal{L}}_r t} & [\ket{\alpha^{M}_I(0)}\bra{\alpha^{-M}_I(0)}]  \\
& = \ket{\alpha^{M}_I(t)}\bra{\alpha^{-M}_I(t)} \ \bra{\alpha^{M'}_I(0)}\alpha^M_I(0)\rangle^{1-e^{-\Gamma t}} \\
& = \ket{\alpha^{M}_I(t)}\bra{\alpha^{-M}_I(t)} \exp[-\frac{1}{2}(|\alpha^M_I(0)|^2 + |\alpha^{-M}_I(0)|^2 \\
& \;\;\;\;\;\;\;\; - 2 \alpha^M_I(0)\alpha^{M'\ast}_I(0)]^{1-e^{-\Gamma t}}  \\
& = \ket{\alpha^{M}_I(t)}\bra{\alpha^{-M}_I(t)} \exp[-2\lambda^2/\omega^2]^{1-e^{-\Gamma t}}\;. \\
\end{split}
\end{equation}

Transforming back to the Schr\"odinger picture, the state written explictly in the qubit basis is:
\begin{equation}
\hat{\rho} (t)=\frac{1}{2} \left(\begin{array}{cc}  \ket{\alpha^{1}(t)}\bra{\alpha^{1}(t)} & e^{ict}e^{\kappa(t)}\ket{\alpha^{1}(t)}\bra{\alpha^{-1}(t)} \\ e^{-ict}e^{\kappa(t)}\ket{\alpha^{-1}(t)}\bra{\alpha^1(t)} & \ket{\alpha^{-1}(t)}\bra{\alpha^{-1}(t)} \end{array}\right)_{Q} ,
\end{equation}
where 
\[
\ket{\alpha^M(t)}=\ket{(1+e^{-\Gamma t/2}e^{i\omega t})\lambda M/2\omega}\;,
\]
the coherently evolved phase is 
\[
c=2mgl/\hbar-\omega_q\;,
\]
and
\[
e^{\kappa(t)}= \exp[-2\lambda^2/\omega^2]^{1-e^{-\Gamma t}}e^{-\Gamma_{\parallel} t}\;.
\]
We seek a form for the joint state after one oscillation period $\tau=2\pi/\omega$. Since $2\pi \Gamma/\omega=Q^{-1}\ll1$, 
we can approximate $\alpha^M(2\pi/\omega)\approx \alpha^M(0)$ and $1-e^{-\Gamma 2\pi/\omega}\approx \Gamma\tau$, so that 
\begin{equation}
e^{\hat{\mathcal{L}}_1 \tau}(\hat{\rho}(0))=\frac{1}{2} \left(\begin{array}{cc}  1 & e^{i\phi}e^{-(\Gamma_{\parallel}+\gamma) \tau} \\ e^{-i\phi}e^{-(\Gamma_{\parallel}+\gamma) 2\pi/\omega} & 1 \end{array}\right)_q \otimes \ket{0}_r\bra{0}\;,
\end{equation}
where the coherent phase is
\begin{equation}
\phi=\frac{2\pi}{\omega} (2mgl/\hbar-\omega_q)\;,
\end{equation}
and the decoherence is governed by the factor
\begin{equation}
\gamma=\frac{2\Gamma l^2}{z_0^2}\;.
\label{eqDecoherenceFactor}
\end{equation}
As expected, the dephasing grows with the square of the cat state separation.

Evolution according to $\hat{\mathcal{L}}_2$ is a map that acts only on the qubit and can be solved for explicitly giving 
\begin{equation}
\begin{array}{lll}
\mathcal{E}^{{\rm (evA)}}(\hat{\rho}(0))&\approx& \left(\begin{array}{cc}  T_1\Gamma_{\perp}(1+N_q-\frac{e^{-\tau/T_1}}{2}) & \frac{1}{2}e^{i\phi}e^{-\gamma \tau}e^{-\tau/T_2} \\ \frac{1}{2}e^{-i\phi}e^{-\gamma \tau}e^{-\tau/T_2} & T_1\Gamma_{\perp}(N_q+\frac{e^{-\tau/T_1}}{2})\end{array}\right)_q\\
&& \otimes \ket{0}_r\bra{0}\;,
\end{array}
\end{equation}
%
%
\section{Transverse trapping}
\label{secTransverseTrapping}

When considering the horizontal movement of the resonator we break the cylindrical symmetry, meaning it is easier to work in Cartesian coordinates. We find the magnetic vector potential and field of the magnetic sphere to be
${\mathbf A} (\mathbf{r}) = \mu_0 {\cal M} V\,(x^2 +y^2+z^2)^{-3/2}\,[-y,x,0]/4 \pi$,
and 
${\mathbf B} (\mathbf{r}) =  \mu_0 {\cal M} V\, (x^2 +y^2+z^2)^{-5/2}\, [-3xz, 3yz, x^2 + y^2 -2z^2] /4 \pi$.
Due to the coordinate system, rather than a circular resonator, we consider a square resonator of width $2w$, and wire radius $a$, and assume it is displaced sideways along the $x-$axis a small amount $\delta x$. We can calculate the flux through the resonator at this position via Eq.~(\ref{eqFluxViaPhiIntegral}), and expand the result in a Taylor series in $\delta x$. To third order we get
\begin{eqnarray}
\Phi(\delta x) &=& \frac{2 \mu_0 w^2 {\cal M} }{\pi (w^2 +z^2)\sqrt{2w^2 +z^2}} \nonumber \\
    && \;\;\; + \frac{\mu_0 w^2 {\cal M}V (5 w^6 - 11 w^4 z^2 - 18 w^2 z^4 - 6 z^6)}{\pi (w^2 +z^2)^3 (2 w^2 +z^2)^{5/2}} \delta x^2 \nonumber \\
    && \;\;\;\;\;\; + O[\delta x^4]\;.
\end{eqnarray}
The zeroth-order term is a constant for motion along the $x-$direction and can be ignored. Using (\ref{eqdIdz}) modified for  $x-$directional motion  we obtain the dependence of the induced current on $\delta x$,
\begin{equation}
I(\delta x) = -\frac{w {\cal M}V (5 w^6 - 11 w^4 z^2 - 18 w^2 z^4 - 6 z^6)}{4 (w^2 +z^2)^3 (2 w^2 +z^2)^{5/2} (\log[2w/a] -0.774)} \delta x^2\;,
\end{equation}
where we have used the fact that self-inductance of a square loop is $L = 2 \mu_0 w (\log[w/a] - 0.774)/\pi$. Using the Lorentz force law as in the previous section, we can integrate the loop current in the presence of the magnetic field and obtain the resulting force.
Renaming the small displacements $\delta x{\sim}\, x$ and similarly for $y,z$ from the equilibrium point $(0,0,z_{\mathrm{eq}})$, we find that to lowest order the $x-$component of this force $F_x = -\beta \, x^3$, $\beta>0$,  and at equilibrium the resonator is transversely trapped in a pure anharmonic potential. As these forces come from a conservative potential we can integrate along paths to obtain the leading terms for the potential of the system
\begin{equation} 
V = \frac{1}{2} m \omega^2 z^2 + \frac{1}{3} \gamma (x^2 + y^2) z + \frac{1}{4} \beta (x^4 + y^4) \, ,
\end{equation}
which describes a type of cross-mode coupling. For parameters described in Appendix \ref{secSymbols} we find
$(m \omega^2 /2, \gamma, \beta)= (1.73 \times 10^{-2}$\,J, $1.98 \times 10^3$\,Jm$^{-1}$, $2.65 \times 10^8$\,Jm$^{-2}$).


%

\end{document}